\documentclass[11pt,a4paper]{article}

\usepackage{jheppub}

\usepackage{epsfig,epsf}
\usepackage{amsmath}
\usepackage{amsthm}
\usepackage{amsfonts}
\usepackage{amssymb}
\usepackage{dsfont}
\usepackage{epstopdf}
\usepackage{multirow}

\usepackage{marvosym}

\usepackage{slashed}

\usepackage[active]{srcltx}
\usepackage{wick}
\def\II{\hbox{{1}\kern-.25em\hbox{l}}}

\newcommand{\der}{\stackrel{\leftrightarrow}{D}}
\newcommand{\derleft}{\stackrel{\leftarrow}{D}}
\newcommand{\derright}{\stackrel{\rightarrow}{D}}

\newcommand{\dhalf}{\frac{d}{2}}

\title{Electroproduction of tensor mesons in QCD}

\author[a]{V. M. Braun}
\author[b,1]{N. Kivel%
\note{On leave of absence from St. Petersburg Nuclear Physics Institute, 188350, Gatchina, Russia}}
\author[a]{M. Strohmaier}
\author[a]{and A. A. Vladimirov}

\affiliation[a]{
   Institut f\"ur Theoretische Physik, Universit\"at
   Regensburg \\ D-93040 Regensburg, Germany}
\affiliation[b]{
Helmholtz Institut Mainz, Johannes Gutenberg-Universit{\"a}t, D-55099 Mainz, Germany}
\emailAdd{vladimir.braun@ur.de}

\emailAdd{kivel@kph.uni-mainz.de}

\emailAdd{matthias.strohmaier@ur.de}

\emailAdd{aleksey.vladimirov@ur.de}

\abstract{
Due to multiple possible polarizations hard exclusive production of tensor mesons by virtual photons or 
in heavy meson decays offers interesting possibilities to study the helicity structure of the underlying 
short-distance process. 
Motivated by the first measurement of the transition form factor $\gamma^*\gamma \to f_2(1270)$ at large momentum 
transfers by the BELLE collaboration we present an improved QCD analysis of this reaction in the framework of 
collinear factorization including contributions of twist-three quark-antiquark-gluon operators 
and an estimate of soft end-point corrections using light-cone sum rules. 
The results appear to be in a very good agreement with the data, in particular 
the predicted scaling behavior is reproduced in all cases.  
       }

\keywords{hard exclusive reactions, factorization, tensor meson}

\begin{document}

\maketitle

\newpage

%%%%%%%%%%%%%%%%%%%%%%%%%%%%%%%%%%%%%%%%%%%%%%%%%%%%%%%%%%%%%%%%%%%%%%%%%%%%%%%%%%%%%%%%%%
\section{Introduction}\label{Sec:Introduction}
%%%%%%%%%%%%%%%%%%%%%%%%%%%%%%%%%%%%%%%%%%%%%%%%%%%%%%%%%%%%%%%%%%%%%%%%%%%%%%%%%%%%%%%%%%

In recent years there has been increasing interest to hard exclusive production of tensor mesons 
$a_2(1320)$, $K^\ast_2(1430)$, $f_2(1270)$ and $f'_2(1520)$
by virtual photons or in heavy meson decays. In particular the possibility of three different polarizations of tensor 
mesons in weak $B$ meson decays can shed light on the helicity structure of the underlying electroweak interactions.
A different symmetry of the wave function and hence a different 
hierarchy of the leading contributions for the tensor mesons as compared to the vector mesons can lead to the 
situations that the color-allowed amplitude is suppressed and becomes comparable to the color-suppressed one.
This feature can give an additional handle on penguin contributions.
The early work was devoted mainly on the identification of the interesting decay modes and their 
basic theoretical description using various factorization techniques at the leading-order and the leading-twist level, 
see e.g.~\cite{Wang:2010ni,Yang:2010qd,Cheng:2010yd,Li:2010ra,Lu:2011jm,Zou:2012sy}.
These studies are to a large extent exploratory. The physics potential of tensor meson production will 
depend on the accuracy of the theoretical description of such processes that can be achieved in QCD.

The recent study~\cite{Masuda:2015yoh} of hard exclusive production of tensor mesons in 
single-tag two-photon processes is an important step forward in this context.  This is  
a ``gold-plated'' reaction where the theoretical formalism can be tested and the relevant nonperturbative functions 
--- tensor meson distribution amplitudes (DAs) --- determined, or at least constrained.   
Our work aims to match this experimental progress with a development of the robust QCD framework for the
study of the transition form factor $\gamma^\ast\gamma \to f_2(1270)$ in collinear factorization. 

This reaction has already attracted some attention. Useful kinematic relations and estimates of the
transition form factors for the mesons built of light and heavy quarks can be found in~\cite{Schuler:1997yw}. 
In Ref.~\cite{Braun:2000cs} it was pointed out that hard exclusive production of $f_2(1270)$ with helicity $\lambda=\pm 2$ 
is dominated by the gluon component in the meson wave function and can be
used to determine gluon admixture in tensor mesons in a theoretically clean manner. 
In Ref.~\cite{Pascalutsa:2012pr} the helicity difference sum rule for the weighted integral of the $\gamma^\ast \gamma$ fusion 
cross section was derived and shown to provide constraints on the transition form factor in question.
A phenomenological model for the tensor meson form factor can also be found in~\cite{Achasov:2015pha}.
A related reaction $\gamma^*\gamma \to \pi\pi$ near the threshold has been discussed in~\cite{Diehl:1998dk,Kivel:1999sd,Diehl:2000uv}.

Theory of the transition form factors goes back to the classical work on hard exclusive 
reactions in QCD~\cite{Chernyak:1977as,Efremov:1979qk,Lepage:1980fj}.
The case of tensor mesons does not bring in complications of principle as compared to the 
pseudoscalar meson transition form factors that have been studied in great detail, 
but the tensor meson case is much less developed on a technical level. Our paper can be viewed as a major update 
of a earlier work~\cite{Braun:2000cs} where the leading contributions to this process have been identified
and calculated at the leading order. The new elements are:
\begin{itemize}
\item{} We introduce twist-three and twist-four DAs and calculate the corresponding contributions to the form factors;
\item{} We calculate meson mass corrections terms in the higher-twist DAs and estimate the leading ``genuine'' three-particle contributions;
\item{} We include the next-to-leading (NLO) corrections and 
        calculate the charm-loop contribution for the helicity amplitude with $\lambda=\pm 2$ 
        taking into account for the $c$-quark mass;
\item{} We estimate quark-gluon coupling constants entering on the higher-twist level using QCD sum rules and 
        the leading-twist gluon couplings using QCD sum rules and, alternatively, 
        from the quarkonium decay $\Upsilon (1S)\rightarrow \gamma\,f_{2}$;  
\item{} We estimate the soft (end-point) correction for the leading, helicity-conserving amplitude.
\end{itemize}   
The main conclusion from our study is that the experimental results on 
the $\gamma^\ast\gamma \to f_2(1270)$ transition form factors reported in Ref.~\cite{Masuda:2015yoh} 
appear to be in a very good agreement with the QCD scaling predictions starting already at moderate 
$Q^2 \simeq 5\,\text{GeV}^2$. This is in contrast to the transition form factors
for pseudoscalar $\pi,\eta,\eta'$ mesons where large scaling violations 
have been observed~\cite{Aubert:2009mc,BABAR:2011ad,Uehara:2012ag}. The absolute normalization for all helicity form factors 
can be reproduced assuming a 10-15\% lower value of the tensor meson coupling to the quark energy-momentum tensor as compared
to the estimates existing in the literature, which is well within the uncertainty.  

The presentation is organized as follows. Section 2 is introductory. It contains the definition of helicity 
amplitudes for the  $\gamma^\ast\gamma\to f_2(1270)$ transition and the necessary kinematic relations.
For the reader's convenience, the relation of our conventions to other definitions existing in the literature 
is explained in Appendix~\ref{app:HelAmp}. Section 3 contains a detailed discussion of the leading-twist and 
higher-twist DAs of the tensor meson, which are the main nonperturbative input in the calculations.
This section contains several new results. The relevant nonperturbative parameters are calculated in Appendix~\ref{app:QCDSR}
using QCD sum rules. In Appendix~\ref{app:Upsilon} we estimate one of the leading-twist gluon couplings from the decay
$\Upsilon (1S)\rightarrow \gamma\,f_{2}$. In Section 4 we calculate the three existing helicity amplitudes in collinear 
factorization, including higher-twist and, partially, radiative corrections. In Section 5 we discuss the power suppressed corrections 
$\sim 1/Q^2$ arising from the end-point regions. We explain how such corrections can be estimated using dispersion relations and duality 
and construct the light-cone sum rule for the largest, helicity conserving amplitude. 
In Section 6 we compare our results to the experimental data~\cite{Masuda:2015yoh} and summarize.    

%%%%%%%%%%%%%%%%%%%%%%%%%%%%%%%%%%%%%%%%%%%%%%%%%%%%%%%%%%%%%%%%%%%%%%%%%%%%%%%%%%%%%%%%%%
\section{$f_2(1270)$ production in two-photon reactions}\label{Sec:Observables}
%%%%%%%%%%%%%%%%%%%%%%%%%%%%%%%%%%%%%%%%%%%%%%%%%%%%%%%%%%%%%%%%%%%%%%%%%%%%%%%%%%%%%%%%%%

We consider the reaction
\begin{equation}
  \gamma^*(q_1)  + \gamma (q_2) \to f_2(P)\,, \qquad q_1^2 =- Q^2\,, \quad q_2^2 = 0\,, \quad P^2 = m^2
\end{equation}
with one real and one virtual photon, $P = q_1 + q_2$. Here and below $m = 1270$~MeV is the meson mass.

The transition amplitude can be related to the matrix element of the time-ordered product of two electromagnetic currents
\begin{eqnarray}
  T_{\mu\nu} = i \int\! d^4 x \,e^{-iq_1 x}\langle f_2(P,\lambda)|T\{j^{\rm em}_\mu(x)j^{\rm em}_\nu(0)\}|0\rangle\,,
\label{eq:Fgamma}
\end{eqnarray} 
where
$$j^{\rm em}_\mu(x) = e_u\bar u(x)\gamma_\mu u(x) +  e_d\bar d(x)\gamma_\mu d(x)+\ldots\,.$$
The correlation function $T_{\mu\nu}$ can be decomposed in contributions of three Lorentz structures
\begin{align}
  T^{\mu\nu} &= T_0^{\mu\nu}+ T_1^{\mu\nu}+ T_2^{\mu\nu}\,,
\end{align}
defined as 
\begin{align}
T_0^{\mu\nu}  &  =e^{(\lambda)\ast}_{\alpha\beta}~\left(  -g^{\mu\nu}_{\bot}\right)
(q_1-q_2)^{\alpha}(q_1-q_2)^{\beta}\frac{m^{2}}{(2q_1q_2)^{2}}T_{0}(Q^{2})\,,
\nonumber\\
T_1^{\mu\nu} &  = e^{(\lambda)\ast}_{\alpha\beta}~\left(  -g_\bot^{\alpha\nu}\right)  (q_1-q_2)^{\beta}
\left[ q_1^\mu -q_2^\mu  \frac{q_1^2}{(q_1q_2)}\right]\frac{m^{2}}{(2q_1q_2)^{2}}T_{1}(Q^{2})\,,
\nonumber\\
T_2^{\mu\nu} &  = e^{(\lambda)\ast}_{\alpha\beta}\left[  g^{\alpha\mu}_{\bot}g^{\beta\nu}_{\bot}
 - \frac{1}{2}g^{\mu\nu}_{\bot}\frac{m^{2}}{(2q_1q_2)^{2}}(q_1-q_2)^{\alpha}(q_1-q_2)^{\beta}\right]  T_{2}(Q^{2})\,. 
\label{defT}
\end{align}
Here
\begin{align}
   g^{\mu\nu}_\bot &= g^{\mu\nu} - \frac{1}{(q_1q_2)}(q_1^\mu q_2^\nu + q_1^\nu q_2^\mu) + \frac{q_1^2}{(q_1q_2)^2} q_2^\mu q_2^\nu\,, &&  2 q_1 q_2 = m^2+Q^2\,.
\end{align}
 The polarization tensor $e^{(\lambda)}_{\alpha\beta}$ is symmetric and traceless, and satisfies the condition 
$e^{(\lambda)}_{\alpha\beta} P^\beta=0$.
Polarization sums can be calculated using 
\begin{equation}\label{polarization}
\sum_\lambda e^{(\lambda)}_{\mu\nu}e^{(\lambda)\ast}_{\rho\sigma}
  = \frac12 M_{\mu\rho} M_{\nu\sigma}+\frac12 M_{\mu\sigma} M_{\nu\rho}  -\frac13 M_{\mu\nu} M_{\rho\sigma}\,,
\end{equation}
where $M_{\mu\nu} = g_{\mu\nu} - P_\mu P_\nu/m^2$ and 
the normalization is such that $e^{(\lambda)}_{\mu\nu}e^{(\lambda')\ast}_{\mu\nu}= \delta_{\lambda\lambda'}$.
The invariant form factors $T_0$, $T_1$ and $T_2$ correspond to the three possible helicity amplitudes 
\begin{eqnarray}\label{Thelicity}
T_0: && \, \gamma^*(\pm 1)+\gamma(\pm 1)\rightarrow f_2(0)\, ,
\nonumber \\
T_1: &&\, \gamma^*(0)+\gamma(\pm 1)\rightarrow f_2(\mp 1)\, ,
\nonumber \\
T_2: &&\, \gamma^*(\pm 1)+\gamma(\mp 1)\rightarrow f_2(\pm 2)\, .
\end{eqnarray}
All three amplitudes (form factors) have mass dimension equal to one 
and scale as $T_k \sim Q^0$ (up to logarithms) in the $Q^2\to \infty$ limit. 
The two-photon decay width of $f_{2}(1270)$ is given by~\cite{Agashe:2014kda}
\begin{equation}
\Gamma[ f_{2}\rightarrow\gamma\gamma]=\frac{\pi\alpha^{2}}{5 m}
\left(  \frac{2}{3}|T_{0}(0)|^{2}+|T_{2}(0)|^{2}\right)
=3.03(40)\,\text{keV}\,,
\label{twophotonwidth}
\end{equation}
where $\alpha \simeq 1/137$ is the electromagnetic coupling constant. 
Assuming that $|T_{2}(0)|\gg|T_{0}(0)|$ we obtain
\begin{equation}
|T_{2}(0)|\simeq\sqrt{ \frac{5 m}{\pi\alpha^{2}}\,\Gamma[f_{2}\rightarrow\gamma\gamma]}=339(22)\,\text{MeV}.
\end{equation}
The relation of our definition of helicity form factors to the other existing in the literature 
definitions is given in Appendix~\ref{app:HelAmp}.

%%%%%%%%%%%%%%%%%%%%%%%%%%%%%%%%%%%%%%%%%%%%%%%%%%%%%%%%%%%%%%%%%%%%%%%%%%%%%%%%%%%%%%%%%%
\section{Distribution amplitudes}\label{Sec:DAs}
%%%%%%%%%%%%%%%%%%%%%%%%%%%%%%%%%%%%%%%%%%%%%%%%%%%%%%%%%%%%%%%%%%%%%%%%%%%%%%%%%%%%%%%%%%

In the standard classification the tensor $J^{PC}=2^{++}$ $SU(3)_f$ nonet is composed of 
$f_2(1270)$, $f'_2(1525)$, $a_2(1320)$ and $K_2^\ast(1430)$. 
Isoscalar tensor states $f_2(1270)$ and $f'_2(1525)$ have a dominant decay 
mode in two pions (or two kaons). The isovector $a_2(1320)$ decays only in three pions
and is more difficult to observe in hard reactions. In the quark model these mesons
are constructed from a constituent quark-antiquark pair in the P-wave and with the total spin 
equal to one. In QCD they can be represented by a set of Fock states in terms of quarks and gluons, 
that further reduce to DAs in the limit of small transverse separations. 

In the exact $SU(3)$-flavor symmetry limit the $f_2(1270)$ meson is
part of a flavor-octet, $f_2 = T_8$, and  $f'_2(1525)$ is a flavor-singlet,
$f_2' = T_1$. However, it is known empirically that the $SU(3)$-breaking 
corrections are large. Since $f_2(1270)$ and $f'_2(1525)$ decay predominantly 
in $\pi\pi$ and $KK$, it follows that they are close to the nonstrange and strange 
flavor eigenstates, respectively, with a small mixing angle, see~\cite{Agashe:2014kda,Li:2000zb}.      
In this paper we assume ideal mixing at a low scale which we take to be
$\mu_0 =1$\,GeV, for definiteness. In other words, we assume that $f_2(1270)$ at this scale is a pure nonstrange isospin 
singlet. This assumption can easily be relaxed when more precise data on the form factors 
become available. 
In what follows the notation $\bar q \ldots q $ refers to the $SU(2)$-flavor-singlet combination
\begin{align}
     \bar q \, q  = \frac{1}{\sqrt{2}} \big[   \bar u\,  u +  \bar d \, d \big]\,,
\end{align} 
where $u$ ans $d$ are the usual ``up'' and ``down'' quark flavors.

Let $n^\mu$ be an arbitrary light-like vector, $n^2=0$, and 
\begin{align}
     p_\mu = P_\mu - \frac12 n_\mu \frac{m^2}{pn}\,, && g_{\mu\nu}^\perp = g_{\mu\nu} - \frac1{pn}\big(n_\mu p_\nu+ n_\nu p_\mu\big)\,.
\end{align}
We define the $f_2$-meson quark-antiquark light-cone DAs as matrix elements of nonlocal light-ray operators~\cite{Braun:2000cs,Cheng:2010hn}
\begin{eqnarray}
 \langle f_2(P,\lambda)| \bar q(z_2n)\gamma_\mu q(z_1n) |0\rangle 
&=&
 f_q m^2 \frac{e^{(\lambda)\ast}_{nn}}{(pn)^2} p_\mu \int_0^1 \!du\, e^{iz_{12}^u (pn)}\,\phi_2(u,\mu)
\notag\\
&&{}+
f_q m^2  \frac{e^{(\lambda)\ast}_{\perp\mu n}}{pn} \int_0^1 \!du\, e^{iz_{12}^u (pn)}\,g_v(u,\mu)
\notag\\
&&{}-\frac12 n_\mu f_q m ^4 \frac{e^{(\lambda)\ast}_{nn}}{(pn)^3} \int_0^1 \!du\, e^{iz_{12}^u (pn)}\,g_4(u,\mu)\,,
\notag\\
 \langle f_2(P,\lambda)| \bar q(z_2n)\gamma_\mu \gamma_5 q(z_1n) |0\rangle 
&=& -i  f_q m^2 \epsilon_{\mu\nu\alpha\beta} \frac{n^\nu p^\alpha}{pn} 
\frac{e^{(\lambda)\ast}_{\beta n}}{pn} \int_0^1 \!du\, e^{iz_{12}^u (pn)}\,g_a(u,\mu)\,, 
\label{DAdef:2pt}
\end{eqnarray}
where 
\begin{align}
 e^{(\lambda)\ast}_{\mu n} \equiv e^{(\lambda)\ast}_{\mu \nu } n^\nu\,, &&   e^{(\lambda)\ast}_{\perp\mu n} \equiv  g^\perp_{\mu\nu}e^{(\lambda)\ast}_{\nu n} = 
e^{(\lambda)\ast}_{\mu n} - p_\mu \frac{e^{(\lambda)\ast}_{nn}}{(pn)}+ \frac12 n_\mu e^{(\lambda)\ast}_{nn} \frac{m^2}{(pn)^2}\,
\end{align}
and we use a shorthand notation
\begin{align}
   z_{12}^u = \bar u z_1 + u z_2\,,\qquad \bar u = 1-u\,.
\end{align}
Note that
\begin{align}
e^{(\lambda)\ast}_{pn} = - \frac12 e^{(\lambda)\ast}_{nn} \frac{m^2}{pn}\,.  
\end{align}
In all expressions light-like Wilson lines between the quark fields are implied. 

The DAs defined in \eqref{DAdef:2pt} satisfy the following symmetry relations: 
\begin{align}
  \phi_2(u) = - \phi_2(\bar u)\,, && g_v(u) = - g_v(\bar u)\,, &&  g_a(u) = + g_a(\bar u)\,, &&  g_4(u) = - g_4(\bar u)\,. 
\end{align}
and are normalized as
\begin{align}
 \int_0^1 \,du\,(2u-1)\phi_2(u) =  \int_0^1 \,du\,(2u-1)g_v(u) =  \int_0^1 \,du\,(2u-1)g_4(u) = 1\,.
\end{align}
The integral of the DA $g_a(u)$ vanishes
\begin{align}
 \int_0^1 \,du\, g_a(u) = 0\,,     
\end{align}
and the first nonzero (second) moment, $\int_0^1 \,du\, (2u-1)^2 g_a(u)$, involves contributions of three-particle operators, see below. 

The coupling $f_q$ is defined as the matrix element of the local operator
\begin{equation}\label{norm1}
\frac12  \langle f_2(P,\lambda)|
\bar q \left[\gamma_\mu i\der_\nu + \gamma_\nu i\der_\mu \right] q |0\rangle =  f_q m^2 e^{(\lambda)\ast}_{\mu\nu}
\end{equation}
where $\der_\mu = \derright_\mu -\derleft_\mu$ is the covariant derivative. This coupling is scale dependent and 
gets mixed with the gluon coupling and the similar coupling for strange quarks. In Appendix~\ref{app:scale}
we summarize the scale dependence of all DA parameters introduced in this Section.
 
The numerical value of $f_q$ has been estimated in the past \cite{Aliev:1981ju,Aliev:1982ab,Cheng:2010hn} 
(see also Appendix~\ref{app:QCDSR}) using the QCD sum rule approach. 
Another possibility is to use the experimental result on the decay width $\Gamma(f_2\to \pi\pi)$ 
and estimate  $f_q$ assuming that the matrix element of the energy-momentum tensor $\langle\pi^+\pi^-|\Theta_{\mu\nu}|0\rangle$ is saturated by 
the tensor meson~\cite{Aliev:1981ju,Aliev:1982ab,Terazawa:1990es,Suzuki:1993zs,Cheng:2010hn}.
These two estimates agree with each other surprisingly well, although this agreement should not be overrated as in both cases
the non-resonant two-pion background is not taken into account. 
We use (cf.~\cite{Cheng:2010hn} and Appendix~\ref{app:QCDSR})  
\begin{align}
       f_q  =  101(10)~\text{MeV}
\end{align}
(at the scale 1 GeV) as the default value for the present study. Note that the positive sign for this coupling is 
a phase convention, whereas the relative signs of the other matrix elements with respect to $f_q$ are physical and 
can be determined by considering suitable correlation functions as explained in Appendix~\ref{app:QCDSR}.  

The operator product expansion (OPE) of quark bilinears close to the light cone $x^2\to 0$ takes the form
\begin{eqnarray}
  \langle f_2(P,\lambda)| \bar q(x)\gamma_\mu q(-x) |0\rangle 
&=&
 f_q m^2 \frac{e^{(\lambda)\ast}_{xx}}{(Px)^2} P_\mu \!\int_0^1 \!\!du\, e^{i(2u-1)(Px)}
\Big[ \phi_2(u) - g_v(u) + \frac14 x^2 m^2 \phi_4(u)\Big]
\notag\\&&{}+
f_q m ^2\frac{e^{(\lambda)\ast}_{\mu x}}{Px} \int_0^1 \!\!du\,  e^{i(2u-1)(Px)}\, g_v(u)
\notag\\&&{}
+\frac12 f_q m^4 x_\mu \frac{e^{(\lambda)\ast}_{xx}}{(Px)^3} 
\int_0^1 \!du\,  e^{i(2u-1)(Px)} \Big[2 g_v(u) - \phi_2(u)-g_4(u)\Big],
\notag\\
\langle f_2(P,\lambda)| \bar q(x)\gamma_\mu \gamma_5 q(-x) |0\rangle 
&=&
-i  f_q m^2 \epsilon_{\mu\nu\alpha\beta} \frac{x^\nu P^\alpha}{Px} \frac{e^{(\lambda)\ast}_{\beta x}}{Px} \int_0^1 \!du\, 
e^{i(2u-1)(Px)}\, g_a(u)\,, 
\end{eqnarray}
where $\phi_4(u)$ is another twist-four two-particle DA that can be expressed in terms of the other functions
using QCD equations of motion (EOM), see below.

In addition we define three-particle twist-three DAs as 
\begin{align}
 g_\perp^{\mu\mu'}\langle f_2(P,\lambda)|\bar q(z_3n) igG_{\mu' n}(z_2n) \slashed{n} q(z_1n) |0\rangle 
&= f_q m^2 (pn) e^{(\lambda)\ast}_{\perp \mu n } \int\mathcal{D}\alpha\,e^{ipn\sum \alpha_k z_k} \Phi_{3}(\alpha)\,,
\notag\\
 g_\perp^{\mu\mu'}\langle f_2(P,\lambda)|\bar q(z_3n) g\widetilde{G}_{\mu' n}(z_2n) \slashed{n}\gamma_5 q(z_1n) |0\rangle 
&= f_q m^2 (pn) e^{(\lambda)\ast}_{\perp \mu n } \int\mathcal{D}\alpha\,e^{ipn\sum \alpha_k z_k} \widetilde{\Phi}_{3}(\alpha)\,.  
\end{align}
The conformal expansion of the three-particle DAs reads~\cite{Ball:1998sk,Braun:2003rp}
\begin{align}
    \Phi_{3}(\alpha) &= 360\alpha_1\alpha_2^2\alpha_3 \Big[\zeta_{3} + \frac12 \omega_{3} (7\alpha_2-3)+\ldots\Big]\,,
\notag\\
    \widetilde{\Phi}_{3}(\alpha) &= 360\alpha_1\alpha_2^2\alpha_3 
\Big[ 0  + \frac12\widetilde{\omega}_{3} (\alpha_1-\alpha_3) +\ldots\Big]\,.
\end{align}

The two-particle DAs $g_a(u)$ and $g_v(u)$ have \emph{collinear}\, twist three and contain contributions
of \emph{geometric}\, twist-two and twist-three operators. The contributions of lower geometric twist
are traditionally referred to as Wandzura-Wilczek (WW) contributions. They can be calculated in the terms of the 
leading-twist DA $\phi_2(u)$ as~\cite{Braun:2000cs,Cheng:2010hn}
\begin{align}
      g^{WW}_v(u) &= \int_0^u\!dv\, \frac{\phi_2(v)}{\bar v} +  \int_{u}^1\!dv\, \frac{\phi_2(v)}{v}\,,
\notag\\
      g^{WW}_a(u) &= \int_0^u\!dv\,\frac{\phi_2(v)}{\bar v}  -   \int_u^1\!dv\, \frac{\phi_2(v)}{v}\,.   
\end{align}
Assuming for simplicity the asymptotic expression for the leading-twist quark DA
\begin{align}
\phi_2^{as}(u) &=  30u(1-u)(2u-1)\,,
\label{phi2as}
\end{align}
one obtains
\begin{align}
      g^{WW}_v(u) &=  3 C^{1/2}_1(2u-1) + 2 C^{1/2}_3(2u-1)\,,
\notag\\
      g^{WW}_a(u) &=  5 C^{1/2}_2(2u-1)\,,
\end{align}
where $C^{1/2}_n(x)$ are Legendre polynomials. The Legendre expansion can be motivated by the properties 
of these DAs under conformal transformations~\cite{Ball:1998sk,Braun:2003rp}.
The ``genuine'' geometric twist-three contributions can be related to the three-particle DAs using EOM:
\begin{align}
g_a(u) &= g^{WW}_a(u) - 10 \zeta_{3} C^{1/2}_2(2u-1) +\frac{15}{8}(\omega_{3}- \widetilde{\omega}_{3}) C^{1/2}_4(2u-1)\,,   
\notag\\
g_v(u)  &= g^{WW}_v(u) - \Big[10 \zeta_3 - \frac{15}{8}(\omega_{3}- \widetilde{\omega}_{3})\Big]C^{1/2}_3(2u-1)\,.  
\end{align}
The twist-three matrix elements  can be estimated using QCD sum rules, see Appendix~\ref{app:QCDSR}.
We obtain  (at the scale 1 GeV)
\begin{align}
 \zeta_3 = 0.15(8)\,, && \omega_3 = -0.2(3)\,, && \widetilde\omega_3  = 0.06(1)\,. 
\label{twist3ref}
\end{align}

The DAs $\phi_4(u)$ and $g_4(u)$ have collinear twist four and receive contributions of the geometric twist-two, -three
and -four operators. The Wandzura-Wilczek-type twist-two contributions assuming the asymptotic expression 
for $\phi_2(u)$~\eqref{phi2as} have the form
\begin{align}
 \phi^{WW}_4(u) &=  100u^2(1-u)^2(2u-1)\,,
\notag\\[2mm]
 g^{WW}_4(u) &=  30u(1-u)(2u-1)\,.
\end{align}
We expect that these contributions are the dominant source of the power-suppressed corrections $\sim 1/Q^2$ 
because of the large mass of the $f_2(1270)$ and will neglect      
``genuine'' geometric twist-three and twist-four contributions.
The derivation of these expressions proceeds similar to the case of the DAs of vector mesons considered 
in~\cite{Ball:1998sk,Ball:1998ff,Ball:2007zt} so that we omit the details.

Finally, the leading-twist gluon DAs of $f_2(1270)$ can be defined as~\cite{Braun:2000cs}
\begin{eqnarray}
g_\perp^{\mu\mu'}g_\perp^{\nu\nu'} \langle f_2(P,\lambda)| G^a_{n\mu'}(z_2n)G^a_{n\nu'}(z_1n')|0\rangle
&=&
 f^T_g \Big[e^{(\lambda)}_{\perp\mu\nu} (pn)^2-\frac12 g^\perp_{\mu\nu} m^2 e^{(\lambda)}_{nn}\Big]  \int_0^1\!du\, e^{iz_{12}^u pn}  \phi_g^T(u)
\notag\\&&{}
- f^S_g  m^2 g^\perp_{\mu\nu}e^{(\lambda)}_{nn} \int_0^1\!du\, e^{iz_{12}^u pn }\phi_g^S(u)\,.
%+ \tilde f^S_g  m^2_T \epsilon^\perp_{\mu\nu}e^{(\lambda)}_{nn} \int_0^1\!du\, e^{iz_{12}^u pn }\tilde \phi_g^S(u)
%\notag
\end{eqnarray}
The distribution amplitudes $\phi_g^T(u)$ and  $\phi_g^S(u)$ are 
both symmetric to the interchange 
of $u\leftrightarrow \bar u$ and describe the momentum fraction distribution
of the two gluons in the $f_2$-meson with the same and the opposite 
helicity, respectively. 
The asymptotic distributions at large scales are equal to 
\begin{equation}\label{phig}
   \phi_g^{T,{\rm as}}(u) = \phi_g^{S,{\rm as}}(u)= 30 u^2(1-u)^2\,. 
\end{equation}  
The normalization constants $f^T_g$ and $f^S_g$ are defined through the matrix element of 
the local two-gluon operator:
\begin{eqnarray}
\langle f_2(P,\lambda)|
G^a_{\alpha\beta}(0)G^a_{\mu\nu}(0)|0\rangle &=&
f^T_g\biggl\{ 
\big[(P_\alpha P_\mu-\frac12 m^2g_{\alpha \mu})\, e^{(\lambda)}_{\beta\nu}-
 (\alpha\leftrightarrow\beta)\big]- (\mu\leftrightarrow\nu)\biggl\}
\nonumber\\
&-&\frac12 f^S_g m^2\biggl\{\big[g_{\alpha \mu}\,e^{(\lambda)}_{\beta\nu}
-  (\alpha\leftrightarrow\beta)\big]- (\mu\leftrightarrow\nu)\biggl\}.
\end{eqnarray}
The coupling $f_g^S$ can be estimated from the radiative decay $\Upsilon(1S)\to\gamma f_2$, see Appendix~\ref{app:Upsilon}.
The result is consistent with the assumption that  $f_g^S$ is very small at hadronic scales and is generated mainly
by the evolution. In the numerical analysis we use the value 
\begin{align}
 f_g^S(1\,\text{GeV}) =  0\,.
\end{align}
The coupling to a helicity-aligned gluon pair, $f_g^T$, is difficult to quantify. 
The calculation of the leading contributions to the relevant correlation functions suggests
that the two couplings, $f_g^S$ and $f_g^T$, have  the same sign, see Appendix~\ref{app:QCDSR}. 
In what follows we use
\begin{align}
 f_g^T(1\,\text{GeV}) \approx 20\,\text{MeV}
\label{fgtref}
\end{align}
as a ballpark estimate.

As already mentioned, all couplings considered here are scale dependent. The relevant 
expressions are collected in Appendix~\ref{app:scale}.

%%%%%%%%%%%%%%%%%%%%%%%%%%%%%%%%%%%%%%%%%%%%%%%%%%%%%%%%%%%%%%%%%%%%%%%%%%%%%%%%%%%%%%%%%%
\section{QCD factorization}\label{Sec:pQCD}
%%%%%%%%%%%%%%%%%%%%%%%%%%%%%%%%%%%%%%%%%%%%%%%%%%%%%%%%%%%%%%%%%%%%%%%%%%%%%%%%%%%%%%%%%%

%%%%%%%%%%%%%%%%%%%%%%%%%%%%%%%%%%%%%%%%%%%%%%%%%%%%%%%%%%%%%%%%%%%%%%%%%%%%%%%%%%%%%%%%%%%%%%%%%%%%%%%%%%%%%%%%%%%%%%
\begin{figure}
\centerline{\includegraphics[width=0.75\linewidth]{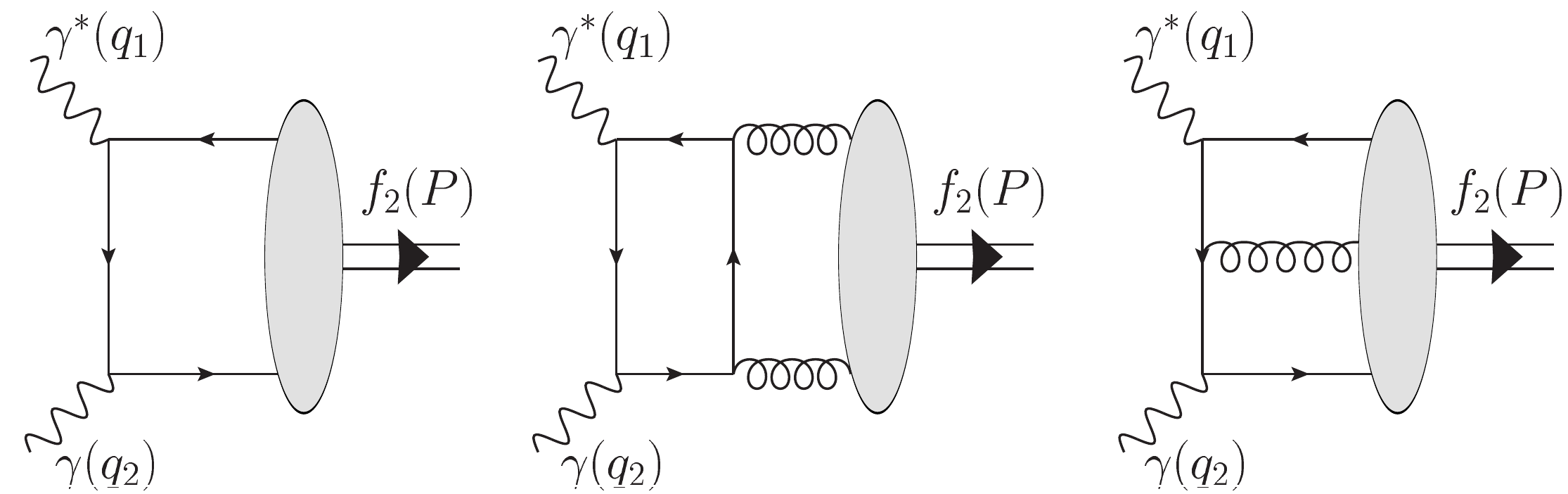}}
\caption{\sf Leading contributions to the transition form factors $\gamma^\ast\gamma\to f_2(1270)$ in QCD. Adding crossing-symmetric diagrams is implied.}
\label{fig:f2gamma}
\end{figure}
%%%%%%%%%%%%%%%%%%%%%%%%%%%%%%%%%%%%%%%%%%%%%%%%%%%%%%%%%%%%%%%%%%%%%%%%%%%%%%%%%%%%%%%%%%%%%%%%%%%%%%%%%%%%%%%%%%%%%%

QCD description of the transition form factors in two-photon reactions is based on the 
analysis of singularities in the product of two electromagnetic currents in \eqref{eq:Fgamma} in the 
limit $(x-y)^2\to 0$. Typical Feynman diagrams contributing to the leading-order accuracy are 
shown in Fig.~\ref{fig:f2gamma}. 

The leading contributions in the $Q^2\to\infty$ limit have been calculated 
already in~\cite{Braun:2000cs}. The form factor $T_0(Q^2)$ is of the leading twist and is dominated by the quark DA.
In this case we include, in addition, NLO perturbative corrections to the leading twist contribution, 
which can be extracted from the corresponding expressions for the two-pion production in~\cite{Kivel:1999sd}.
We also include the twist-four meson-mass correction $m^2/Q^2$ which is a new result.

The form factor $T_1(Q^2)$ is of twist-three. It receives the Wandzura-Wilczek-type contributions calculated in~\cite{Braun:2000cs} 
and the ``genuine'' twist-three contributions of three-particle quark-antiquark gluon DAs (new result).

As already noticed in~\cite{Braun:2000cs}, the  $T_2(Q^2)$ form factor is rather peculiar: The leading contribution at $Q^2\to \infty$ 
comes in this case from the two-gluon DA with aligned helicity that we refer to as gluon transversity DA. 
However, this contribution is suppressed by the factor $\alpha_s/\pi \sim 0.1$ which is the standard perturbation 
theory factor for an extra loop, and also the two-gluon coupling to a ``conventional'' quark-antiquark meson is
expected to be small as compared to the quark-antiquark coupling. By this reason the true QCD asymptotics
for this form factor may be postponed to very large momentum transfers that are out of reach on the existing 
experimental facilities. The result for $T_2(Q^2)$ given below includes the leading term and the 
Wandzura-Wilczek-type higher-twist power correction that does not involve such small factors.      
We also calculate and add the leading-twist $c$-quark contribution. 

With these new additions, the expressions for the form factors are 
\begin{eqnarray}
 T_0&=& \langle f_q\rangle
\int_0^1\! \frac{du}{\bar u}\,\biggl[1 + \frac{\alpha_s}{4\pi}\mathbb{C}_q(u)\biggr]\phi_2(u)
- \frac{\alpha_s}{4\pi} \frac23 f_g^S \int_0^1\!du\,\mathbb{C}^S_g(u)\,\phi_g^S(u)
\notag\\&&{}
+\frac{2 m^2}{Q^2}\langle f_q\rangle \int_0^1\! \frac{du}{\bar u}\left[ u\ln u\,\phi_2(u) - \frac{1}{8\bar u}\phi_4(u)\right],
\label{QCDT0}\\
T_1&=& 2 \langle f_q\rangle \int_0^1 \frac{du}{\bar u}\Big[g_v(u)-g_a(u)\Big]
\nonumber\\&=&
 4 \langle f_q\rangle \int_0^1 \frac{du}{\bar u} \ln (u) \,\phi_2(u)
+  2 \langle f_q\rangle \int\!\mathcal{D\alpha}\,
 \mathbb{C}_\Phi(\alpha) \big[\Phi_3(\alpha) + \widetilde{\Phi}_3(\alpha)\big],
\label{QCDT1}\\
T_2&=&
 \frac{4 m^2}{Q^2}\langle f_q\rangle \int_0^1\!du\,\ln u\,g_v(u)
+\frac{\alpha_s}{\pi}f_g^T\int_0^1 \frac{du}{\bar u}\left[\frac23 + \frac49  \mathbb{C}_c(u)\right] \phi_g^T(u)\,,
\label{QCDT2}
\end{eqnarray}
where the notation $\langle f_q\rangle$ stands for the sum of the light quark couplings weighted with the electromagnetic charges
\begin{align}
  \langle f_q\rangle = \frac49 f_u(\mu) + \frac19 f_d(\mu) + \frac19 f_s(\mu) = \frac{5\sqrt{2}}{18} f_q(\mu) + \frac19 f_s(\mu)\,. 
\end{align} 
The coefficient function of the three-particle DAs to $T_1$ is given by 
\begin{align}
    \mathbb{C}_\Phi(\alpha) &= \frac{1}{\alpha_2}
\left[\frac{1}{\alpha_1\bar\alpha_1} + 
\frac{1}{\alpha_2}\left(\frac{\ln\alpha_1}{\bar\alpha_1}- \frac{\ln\bar \alpha_3}{\alpha_3}\right) 
+\frac{\ln\alpha_1}{\bar\alpha_1^2}\right],
\end{align} 
and the NLO quark and gluon coefficient functions for $T_0$ read~\cite{Kivel:1999sd}
\begin{align}
\mathbb{C}_q(u) = C_F\biggl[ \ln^2 \bar u + 3 \ln u  - 9 \biggr],
&&
\mathbb{C}^S_g(u) =  \frac{2 \ln u}{u\bar u^2}\biggl[
  u \ln u - 2 u    - 2
\biggr].
\end{align}
The $c$-quark contribution to the transversity gluon distribution (this is a new result) is given by 
\begin{eqnarray}
 \mathbb{C}_c(u) &=& 1
 +\frac{2 m^2_c}{Q^2}\biggl[-\frac{\beta}{u\bar u}\ln\left(\frac{\beta+1}{\beta-1}\right)+\frac{\beta_u}{\bar u}\ln\left(\frac{\beta_u+1}{\beta_u-1}\right)
+\frac{\beta_{\bar u}}{u}\ln\left(\frac{\beta_{\bar u}+1}{\beta_{\bar u}-1}\right)
\\&&\nonumber~~~~~~~~~~~+\frac{1}{u\bar u}\left(\frac{1}{2}+\frac{m^2_c}{Q^2}\right)
\left(\ln^2\left(\frac{\beta+1}{\beta-1}\right)-\ln^2\left(\frac{\beta_u+1}{\beta_u-1}\right)-\ln^2\left(\frac{\beta_{\bar u}+1}{\beta_{\bar u}-1}\right)\right)
\biggr],
\end{eqnarray}
where
\begin{align}
\beta_u=\sqrt{1+\frac{4m_c^2}{u Q^2}}\,, && \beta \equiv \beta_1.
\end{align}
Here $m_c \simeq 1.4\,\text{GeV}$ is the $c$-quark mass.

We have checked the electromagnetic gauge invariance of our results by explicit calculation.
Note that electromagnetic Ward identities relate the contributions of three-particle DAs 
(in the last diagram in Fig.~\ref{fig:f2gamma}) to the contributions corresponding to gluon emission from the external quark legs in the 
hard scattering amplitude that are encoded in the ``genuine'' twist-three contributions to the two-particle DAs. 
Thus it is not surprising that the twist-three form factor $T_1(Q^2)$ can be written in two equivalent representations
as in \eqref{QCDT1}: either contributions of the three-particle DAs can be eliminated in favor the two particle ones,
or, vice verse, the ``genuine'' twist-three contributions to the two-particle DAs can be rewritten in terms of the three-particle DAs.    

Evaluating \eqref{QCDT0}, \eqref{QCDT1}, \eqref{QCDT2} on the asymptotic expressions for all DAs that are 
collected in the previous Section we obtain 
\begin{eqnarray}
T_0 &=&  5 \left(1 - \frac{\alpha_s}{27\pi}\right) \langle f_q\rangle - \frac{215}{27} \frac{\alpha_s}{\pi} f_g^S 
 -  5 \frac{m^2}{Q^2} \langle f_q\rangle\,,
\\
T_1 &=& \frac{10}{3}\langle f_q\rangle\Big[ 1+ 4 \zeta_3+\frac{9}{16}(\omega_3-\tilde \omega_3)\Big],
\\
T_2&=&
\frac{10}{3} \frac{m^2}{Q^2}\langle f_q\rangle \Big[2- \zeta_3+\frac{3}{16}(\omega_3-\tilde \omega_3)\Big]~+~
\frac52 \frac{\alpha_s}{\pi}f_g^T\Big[\frac23 + \frac49\lambda(m_c^2/Q^2)\Big]\,,
\end{eqnarray}
where all nonperturbative parameters and the QCD coupling have to be taken at the hard scale $\mu \propto Q$. The function $\lambda(m_c^2/Q^2)$ 
takes into account suppression of the charm quark contribution in comparison to the light flavors; it is given by
\begin{align}
 \lambda(x) &= 1-30x-72 x^2 + 24 x (1+3 x)\widehat{\beta} \ln\left(\frac{\widehat{\beta}+1}{\widehat{\beta}-1}\right)-
6 x \left(1+6 x+12 x^2 \right)\ln^2\left(\frac{\widehat{\beta}+1}{\widehat{\beta}-1}\right),
\end{align}
where $\widehat{\beta} = \sqrt{1+4 x}$. The normalization is such that $\lambda(0)=1$. Note that $\lambda(0.1) \simeq 0.091$
so that the $c$-quark contribution at $Q^2\sim 20\,\text{GeV}^2$ is still suppressed by an order of magnitude as compared to 
the contributions of $u$, $d$, $s$ quarks.

The expressions for the helicity form factors collected in this Section present our main result.

%%%%%%%%%%%%%%%%%%%%%%%%%%%%%%%%%%%%%%%%%%%%%%%%%%%%%%%%%%%%%%%%%%%%%%%%%%%%%%%%%%%%%%%%%%%%%%%%%%%%%%%%%%%%%%%%%%%%%%
\section{Soft (end-point) contributions}\label{Sec:Soft}
%%%%%%%%%%%%%%%%%%%%%%%%%%%%%%%%%%%%%%%%%%%%%%%%%%%%%%%%%%%%%%%%%%%%%%%%%%%%%%%%%%%%%%%%%%%%%%%%%%%%%%%%%%%%%%%%%%%%%%

Transition form factors with one real photon receive power corrections $\sim 1/Q^2$ coming from the region of 
large separation $(x-y)^2 \sim 1/\Lambda_{\rm QCD}^2$ between the electromagnetic currents in \eqref{eq:Fgamma}.
Such contributions are missing in the OPE and involve overlap integrals of the 
nonperturbative light-front wave functions at large transverse separations between the constituents and 
cannot be calculated in terms of DAs. They are revealed, nevertheless, 
as end-point divergences in the momentum fraction integrals in the framework of QCD factorization 
if one tries to extend it beyond the leading power accuracy.   
Such divergences are a clear indication that some extra contributions have to be added.

The technique that we adopt in what follows has been suggested originally~\cite{Khodjamirian:1997tk} for 
the $\gamma^*\gamma \to \pi^0$, see~\cite{Agaev:2010aq,Bakulev:2011rp} for two recent state-of-the-art analysis. In this section we 
demonstrate how the same approach can be applied to the production of tensor mesons (cf.~\cite{Kivel:2000rq}). 
To this end we consider the simplest example: the form factor $T_0$ to the leading-order accuracy, leaving the 
NLO corrections to $T_0$ and the other two form factors, $T_1$ and $T_2$, for a future study.  
Our presentation follows closely the work~\cite{Agaev:2010aq} where further details and generalizations can be found. 
  
The idea is to calculate the transition form factor for two large virtualities 
\begin{align*}
  q_1^2 = -Q^2\,, && q_2^2 = -q^2\,, && Q^2 \gg q^2\,,
\end{align*}
using collinear factorization or, equivalently, OPE, 
and perform analytic continuation to the real photon limit
$q^2=0$ using dispersion relations. In this way explicit evaluation of contributions
of the end-point regions is avoided (since they do not contribute for sufficiently large $q^2$) and effectively 
replaced by certain assumptions on the physical spectral density in 
the $q^2$-channel.    

For our purposes it is sufficient to assume that the second photon is transversely polarized.  
Then there are no new Lorentz structures and the only difference is that the form factors depend on two virtualities.
The starting point is that the form factor
\begin{align}
    \widehat{T}_0(Q^2,q^2) = \frac{1}{(2q_1 q_2)^2}\,{T}_0(Q^2,q^2)\,, 
&& T_0(Q^2) \equiv (m^2+Q^2)^2\,\widehat{T}_0(Q^2,q^2=0)\,,  
\end{align}
satisfies an unsubtracted dispersion relation in the variable $q^2$ for fixed $Q^2$.
Separating the contribution of the lowest-lying vector mesons $\rho,\omega$ one can write
\begin{eqnarray}
\widehat{T}_0^{\gamma^*\gamma^* \to f_2}(Q^2,q^2) &=& \frac{\sqrt{2}f_\rho \widehat{T}_0^{\gamma^*\rho \to f_2}(Q^2)}{m^2_\rho+q^2}
 +
\frac{1}{\pi}\int_{s_0}^\infty ds\,\frac{\mathrm{Im}\, \widehat{T}_0^{\gamma^*\gamma^* \to f_2}(Q^2,-s)}{s+q^2},
\label{eq:DR}
\end{eqnarray}
where $s_0$ is a certain effective threshold. 
Here we assumed that the $\rho$ and $\omega$ contributions
are combined in one resonance term  and the zero-width approximation is adopted; 
$f_\rho\sim 200$~MeV is the usual vector meson decay constant. Since there are no massless states, 
the real photon limit is recovered by the simple substitution $q^2\to 0$ in this equation.

If both virtualities are large, $Q^2, q^2 \gg \Lambda^2_{\rm QCD}$, 
the same form factor can be calculated using  OPE.
Assume this calculation is done to some accuracy.
The result is an analytic function that satisfies a similar dispersion relation
\begin{equation}
\widehat{T}_{0,{\rm OPE}}^{\gamma^*\gamma^* \to f_2}(Q^2,q^2) = \frac{1}{\pi}\int_{0}^\infty ds\,\frac{\mathrm{Im}\, \widehat{T}_{0,{\rm OPE}}^{\gamma^*\gamma^* \to f_2}(Q^2,-s)}{s+q^2}.
\label{eq:DROPE}
\end{equation}
The basic assumption of the method is that the physical spectral density
above the threshold $s_0$ coincides (if integrated with a smooth test function) with the spectral density calculated in OPE, 
in the very similar way as the total cross section of $e^+e^-$-annihilation above the resonance region 
coincides with the QCD prediction, 
\begin{equation}
\widehat{T}_{0,{\rm OPE}}^{\gamma^*\gamma^* \to f_2}(Q^2,-s) \simeq \widehat{T}_{0}^{\gamma^*\gamma^* \to f_2}(Q^2,-s)\,, \qquad s>s_0\,.
\label{eq:duality}
\end{equation}
We expect that OPE becomes exact as $q^2\to \infty$ so that in this limit the calculation has to reproduce the ``true'' form factor.
Equating the two representations in Eqs.~(\ref{eq:DR}), (\ref{eq:DROPE}) at $q^2\to \infty$ and subtracting the contributions of $s>s_0$ from 
the both sides one obtains
\begin{eqnarray}
  \sqrt{2}f_\rho \widehat{T}_0^{\gamma^*\rho \to f_2}(Q^2) = \frac{1}{\pi}\int_{0}^{s_0}\!\!ds\,
\mathrm{Im}\, \widehat{T}_{0,{\rm OPE}}^{\gamma^*\gamma^* \to f_2}(Q^2,-s)\,.
\end{eqnarray}
This relation explains why $s_0$ is usually referred to as the interval of duality (in the vector channel).
The (perturbative) QCD spectral density $\mathrm{Im}\, \widehat{T}_{0,{\rm OPE}}^{\gamma^*\gamma^* \to f_2}(Q^2,-s)$ 
is a smooth function, very different from the physical spectral 
density $\mathrm{Im}\,\widehat{T}_0^{\gamma^*\gamma^* \to f_2}(Q^2,-s) \sim \delta(s-m_\rho^2)$.
Nevertheless, their integrals over a certain region of energies coincide. In this sense QCD
description of correlation functions in the terms of quark and gluons is dual to the description in the terms of
hadronic states.

In practical applications one uses a certain trick~\cite{SVZ} which allows to reduce the sensitivity on the duality
assumption in \eqref{eq:duality} and simultaneously suppress contributions of higher twists in the OPE. 
This is done going over to the Borel representation $1/(s+q^2)\to \exp[-s/M^2]$
the net effect being the appearance of an additional weight factor under the integral that 
suppresses the large invariant mass region:
\begin{eqnarray}
  \sqrt{2}f_\rho \widehat{T}_0^{\gamma^*\rho \to f_2}(Q^2) &=& \frac{1}{\pi}\int_{0}^{s_0}ds\, e^{-(s-m^2_\rho)/M^2}\,
\mathrm{Im}\,\widehat{T}_{0,{\rm OPE}}^{\gamma^*\gamma^* \to f_2}(Q^2,-s)\,.
\label{eq:Frhogamma}
\end{eqnarray}
Varying the Borel parameter $M^2$ within a certain window, usually 1--2~GeV$^2$ one can test 
sensitivity of the results to the particular model of the spectral density.

With this refinement, substituting Eq.~(\ref{eq:Frhogamma}) in (\ref{eq:DR}) and using
Eq.~(\ref{eq:duality}) one obtains for $q^2\to 0$ (cf.~\cite{Khodjamirian:1997tk})
\begin{align}
\widehat{T}_{0,{\rm LCSR}}^{\gamma^*\gamma \to f_2}(Q^2) &=
\frac{1}{\pi}\!\int_{0}^{s_0}\! \frac{ds}{m_\rho^2}\,e^{(m^2_\rho-s)/M^2}\, 
\mathrm{Im} \widehat{T}_{0,{\rm OPE}}^{\gamma^*\gamma^* \to f_2}(Q^2,-s)
+
\frac{1}{\pi}\int_{s_0}^\infty \frac{ds}{s} \mathrm{Im}\,
\widehat{T}_{0,{\rm OPE}}^{\gamma^*\gamma^* \to f_2}(Q^2,-s)\,. 
\label{eq:LCSR1}
\end{align}
The abbreviation LCSR stands for the Light-Cone Sum Rules~\cite{Balitsky:1989ry}, as this approach is usually 
referred to. 

Adding and subtracting the contribution of $s<s_0$ in the second term%
\footnote{Such a rewriting is not always possible as the separation of the OPE result and the soft correction 
 can suffer from end-point divergences, see~\cite{Agaev:2010aq}.}, 
 one can rewrite the result as
\begin{eqnarray}
\widehat{T}_{0,{\rm LCSR}}^{\gamma^*\gamma \to f_2}(Q^2) &=& 
\widehat{T}_{0,{\rm OPE}}^{\gamma^*\gamma \to f_2}(Q^2) + \widehat{T}_{0,{\rm soft}}^{\gamma^*\gamma \to f_2}(Q^2)\,, 
\end{eqnarray}
where the first term is the original OPE expression which is possible but not necessary to write in the dispersion
representation, and the second term is the correction of interest:
\begin{eqnarray}
\widehat{T}_{0,{\rm soft}}^{\gamma^*\gamma \to f_2}(Q^2) &=&
\frac{1}{\pi}\!\int_{0}^{s_0}\!\frac{ds}{s} \left[\frac{s}{m_\rho^2}\,e^{(m^2_\rho-s)/M^2}\,-1\right] 
\mathrm{Im}\, \widehat{T}_{0,{\rm QCD}}^{\gamma^*\gamma^* \to f_2}(Q^2,-s)\,.
\label{eq:LCSR2}
\end{eqnarray} 
An attractive feature of this technique is that one does not need to evaluate the nonperturbative wave function overlap contributions 
explicitly: They are taken into account effectively via the modification of the spectral density.

As an illustration, consider the leading-twist QCD result at the leading order in strong coupling:
\begin{eqnarray}
 \widehat{T}_0(Q^2,q^2) &=&  \langle f_q\rangle  \int_{0}^{1}\! du\,\frac{\widetilde\phi_2(u)}{[\bar u Q^2 + u\bar u m^2 + u q^2]^2},
\end{eqnarray}
where 
\begin{align}
  \widetilde\phi_2(u) = - \int_0^u\!dv\, \phi_2(v)\,, && \widetilde\phi^{\rm as}_2(u) = 15 u^2\bar u^2\,. 
\end{align}
This expression can easily be brought to the form of a dispersion relation changing variables
$ u \to s = \frac{\bar u}{u} Q^2 + \bar u m^2$ and integrating by parts.
In this way one obtains after some rewriting,
\begin{eqnarray}
  \widehat{T}_{0,{\rm soft}}^{\gamma^*\gamma \to f_2}(Q^2) &=& 
  -\langle f_q\rangle
 \int_{u_0}^{1}\! du\,\widehat\phi_2(u)\left[\frac{1}{[\bar u Q^2 + u\bar u m^2]^2}- \frac{e^{(m^2_\rho-s)/M^2}}{m_\rho^2 u^2 M^2}\right]  
\notag\\&&{}
+ \langle f_q\rangle\left[\frac{1}{m_\rho^2}\,e^{(m^2_\rho-s_0)/M^2}\,-\frac{1}{s_0}\right] \frac{\widehat\phi_2(u_0)}{u_0^2 m^2 + Q^2},
\end{eqnarray}
where 
\begin{align}
 u_0 = \frac{1}{2m^2} \left[\sqrt{(Q^2+s_0-m^2)^2 + 4 m^2 Q^2 }- (Q^2+s_0-m^2)\right],
\end{align}
and, for comparison, to the same accuracy 
\begin{align}
   \widehat{T}_{0,{\rm OPE}}^{\gamma^*\gamma \to f_2}(Q^2) &=    
\langle f_q\rangle  \int_{0}^{1}\! du\,\frac{\widehat\phi_2(u)}{[\bar u Q^2 + u\bar u m^2]^2}.
\end{align}
Note that $u_0\to 1$ as $Q^2 \to \infty$ so that the integration region shrinks to the 
end-point $u\to 1$ and the correction is power suppressed $\sim 1/Q^2$ in this limit, as expected.  
Numerical results are presented in the next Section.

%%%%%%%%%%%%%%%%%%%%%%%%%%%%%%%%%%%%%%%%%%%%%%%%%%%%%%%%%%%%%%%%%%%%%%%%%%%%%%%%%%%%%%%%%%
\section{Results and discussion}\label{Sec:Results}
%%%%%%%%%%%%%%%%%%%%%%%%%%%%%%%%%%%%%%%%%%%%%%%%%%%%%%%%%%%%%%%%%%%%%%%%%%%%%%%%%%%%%%%%%%

%%%%%%%%%%%%%%%%%%%%%%%%%%%%%%%%%%%%%%%%%%%%%%%%%%%%%%%%%%%%%%%%%%%%%%%%%%%%%%%%%%%%%%%%%%%%%%%%%%%%%%%%%%%%%%%%%%%%%%%
\begin{figure}[ht]
\includegraphics[width=0.99\linewidth, trim=1.2cm 11.0cm 1.2cm 10.5cm, clip = true]{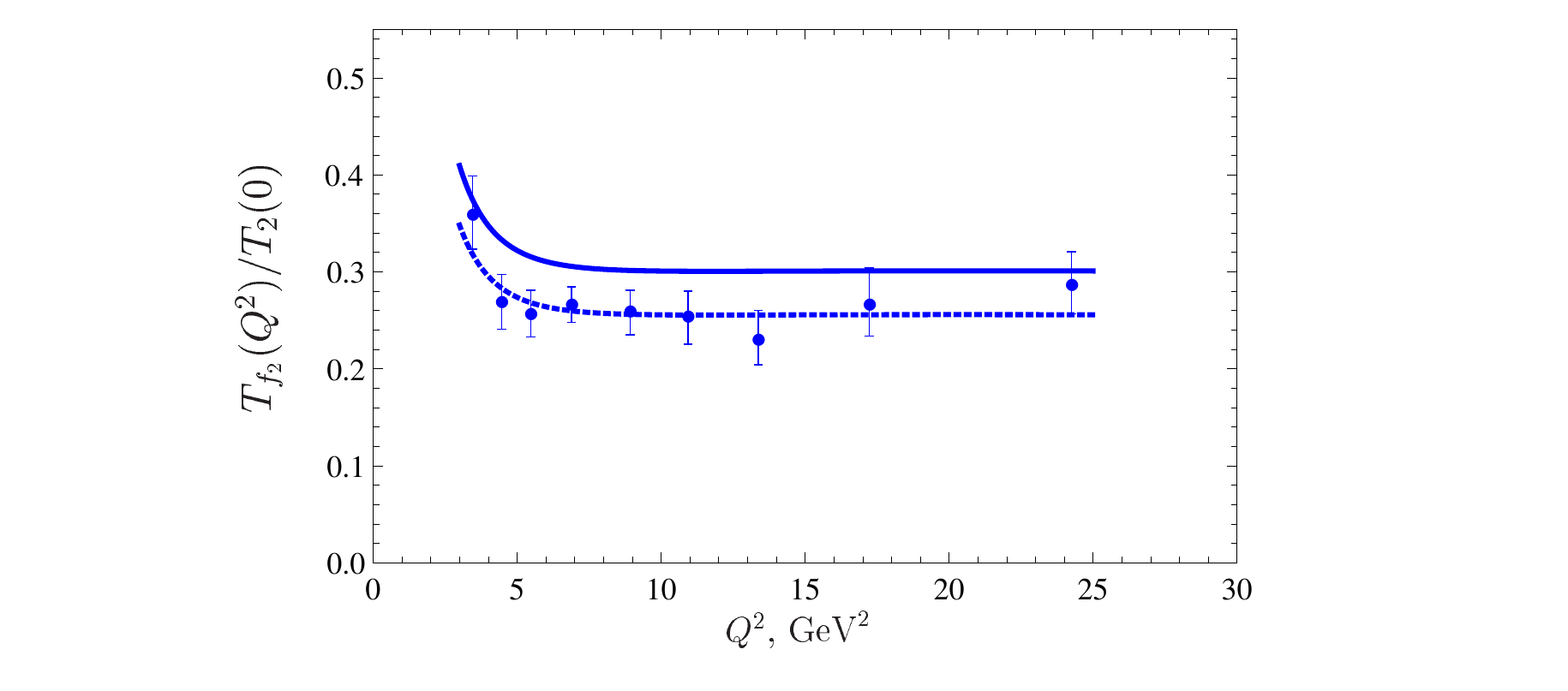}
\caption{\sf The effective form factor summed over polarizations normalized to $T_{2}(0)= 339\, \text{MeV}$.
The calculation using default values of the nonperturbative parameters is shown by the sold curve. 
The same calculation with the quark coupling $f_q$ reduced by 15\% is shown by short dashes.
The experimental data are taken from Ref.~\cite{Masuda:2015yoh}. Only statistical errors are shown.
}

\label{fig:Teff}
\end{figure}
%%%%%%%%%%%%%%%%%%%%%%%%%%%%%%%%%%%%%%%%%%%%%%%%%%%%%%%%%%%%%%%%%%%%%%%%%%%%%%%%%%%%%%%%%%%%%%%%%%%%%%%%%%%%%%%%%%%%%%%

The effective form factor averaged over polarizations 
\begin{align}
       T_{f_2}(Q^2) &= 
\sqrt{\frac23 \left\vert\frac{T_{0}(Q^{2})}{T_{2}(0)}\right\vert^2 
     + \frac{Q^2m^2}{(m^2+Q^2)^2} \left\vert\frac{T_{1}(Q^{2})}{T_{2}(0)}\right\vert^2 
     + \left\vert\frac{T_{2}(Q^{2})}{T_{2}(0)}\right\vert^2}\,,    
\end{align}
is calculated using default values of the nonperturbative parameters and compared with the 
experimental data~\cite{Masuda:2015yoh} in Fig.~\ref{fig:Teff}.
We observe a perfect scaling behavior for $Q^2 \gtrsim 4-5\,\text{GeV}^2$ as predicted by QCD,
whereas the normalization is slightly off --- about $1-1.5\sigma$ if systematic errors in the data are taken into account.
 This difference can easily be compensated by a $10-15\%$ decrease of the value of the quark coupling $f_q$  which serves as an 
overall normalization factor in the calculation, or, alternatively, by a moderate deviation of the leading twist DA $\phi_2(u)$
from its asymptotic form.
For illustration we show in the same Figure by short dashes the 
result of the QCD calculation with $f_q = 85\,\text{MeV}$ at the scale 1 GeV.

Such a $10-15\%$ smaller value of $f_q$ is certainly possible and does not contradict the existing estimates which are not very 
reliable. A more precise value can eventually be obtained from lattice QCD, 
however, this calculation is rather complicated and will take time. It would be very interesting 
to measure the time-like transition form factor $e^+e^-\to f_2(1270)+\gamma$ at large 
virtualities $q^2 \sim 100\,\text{GeV}^2$ (cf.~\cite{Aubert:2006cy}) 
where the nonperturbative uncertainties are considerably reduced. This would give a direct measurement of the $f_q$-coupling.   

%%%%%%%%%%%%%%%%%%%%%%%%%%%%%%%%%%%%%%%%%%%%%%%%%%%%%%%%%%%%%%%%%%%%%%%%%%%%%%%%%%%%%%%%%%%%%%%%%%%%%%%%%%%%%%%%%%%%%%
\begin{figure}
\begin{center}
\includegraphics[width=0.99\linewidth, trim=1.2cm 3.5cm 1.2cm 3.5cm, clip = true]{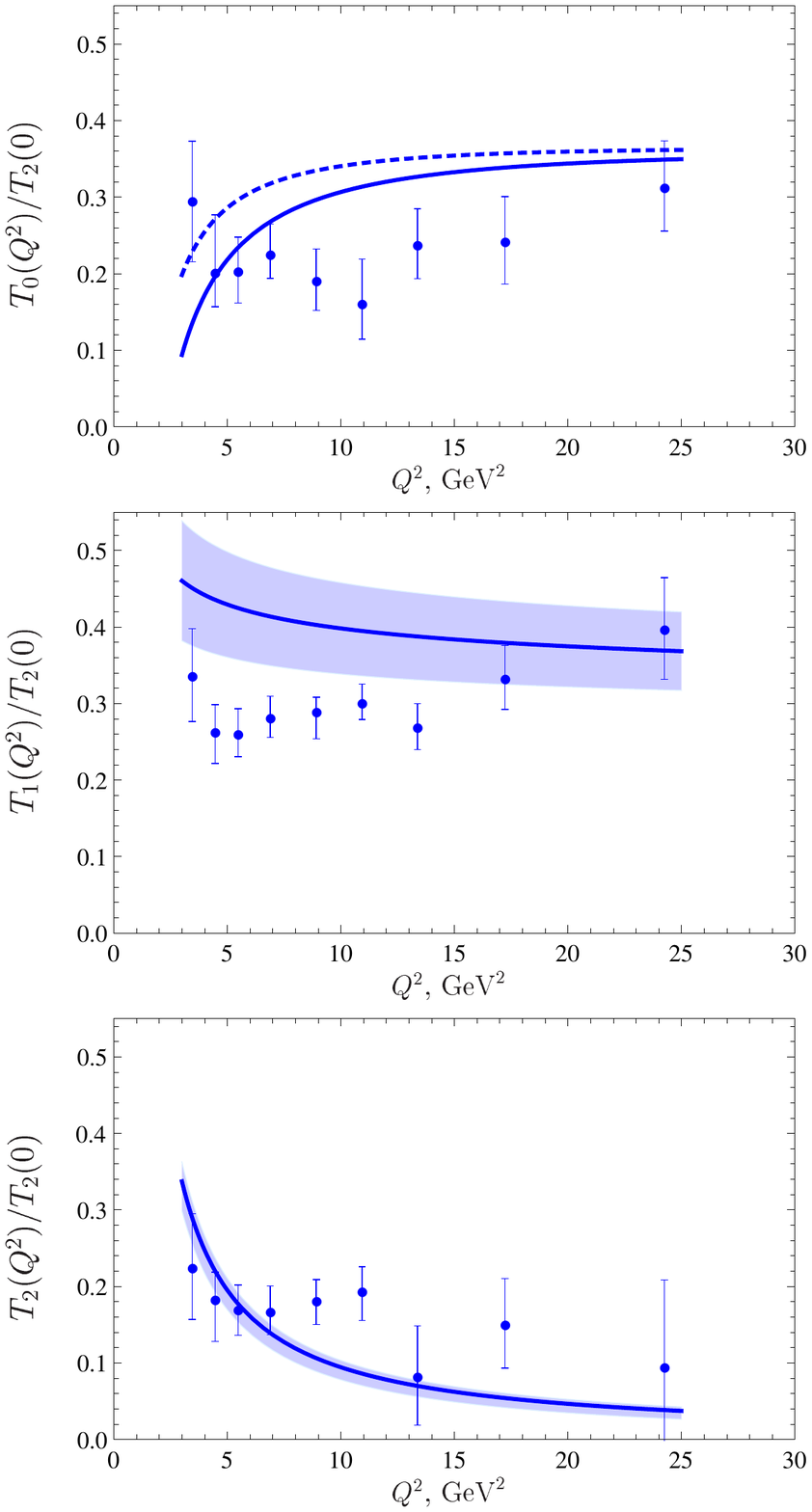}
\end{center}
\vspace*{-0.5cm}
\caption{\sf The form factors $T_0(Q^2)$, $T_1(Q^2)$, $T_2(Q^2)$ (from top to bottom) normalized to $T_2(0)= 339\, \text{MeV}$.
 The result for  $T_0(Q^2)$ shown by the solid line includes the estimate of soft end-point contributions using light-cone 
 sum rules. The result of a pure pQCD calculation is shown by dashes.
 The error band for $T_1(Q^2)$ (shaded area) corresponds to variation of the twist-three 
 parameters in the range specified in \eqref{twist3ref}, whereas for $T_2(Q^2)$ we also include variation 
 of the tensor gluon coupling $f_g^T$ in the range $\pm 50\,\text{MeV}$.
The experimental data are taken from Ref.~\cite{Masuda:2015yoh}. Only statistical errors are shown.}
\label{fig:QCDvsEXP}
\end{figure}
%%%%%%%%%%%%%%%%%%%%%%%%%%%%%%%%%%%%%%%%%%%%%%%%%%%%%%%%%%%%%%%%%%%%%%%%%%%%%%%%%%%%%%%%%%%%%%%%%%%%%%%%%%%%%%%%%%%%%%

Our results for the
helicity-separated form factors  $T_0(Q^2)$, $T_1(Q^2)$, $T_2(Q^2)$ are compared with the 
experimental data~\cite{Masuda:2015yoh} in Fig.~\ref{fig:QCDvsEXP}. All three form factors are
described rather well, the QCD calculation being slightly above the data as we have already seen 
for the helicity-averaged  form factor in Fig.~\ref{fig:Teff}. Note that our result for $T_1(Q^2)$ only 
includes the leading-power contribution at large $Q^2$ in contrast to $T_0(Q^2)$ and $T_2(Q^2)$ where we
also calculated the $1/Q^2$ correction. Terms $\sim 1/Q^2$ in $T_1(Q^2)$ correspond to collinear-twist-five
and soft contributions and are more difficult to estimate. They should be expected, however, 
to be negative and of the same order of magnitude as for $T_2(Q^2)$ so that the increase of the QCD curve 
for $T_1(Q^2)$ in Fig.~\ref{fig:QCDvsEXP} at smaller $Q^2$ will almost certainly be compensated by power corrections and 
is not a reason for concern.  As expected, $T_1(Q^2)$ is also more sensitive to the twist-three quark-antiquark-gluon
contributions as compared to the other two form factors, 
and the uncertainties in the corresponding parameters are not negligible, they are shown by the shaded area.  

As discussed in~\cite{Braun:2000cs}, the form factor $T_2(Q^2)$ at asymptotically large $Q^2$ is dominated by 
the two-gluon contribution with aligned helicity that we refer to as gluon transversity DA. This contribution is 
suppressed, however, by the factor $\alpha_s/\pi \sim 0.1$ which is the standard penalty for an extra loop. 
Also the two-gluon coupling to a ``conventional'' quark-antiquark meson is
unlikely to be large as compared to the quark-antiquark coupling. 
By this reason, $T_2(Q^2)$ at realistic $Q^2$ is still dominated by the
Wandzura-Wilczek-type higher-twist power correction that does not involve such small factors: 
The shaded area in the plot for  $T_2(Q^2)$ includes variation of the tensor gluon coupling $f_g^T$ in a rather broad range, 
$\pm 50\,\text{MeV}$, but the effect is barely visible. 
Our result does not mean that measurements of the $T_2$ form factor at large $Q^2$ are not interesting. On the contrary,
a broad resonance structure in the two-pion channel with a scaling behavior $T_0\sim Q^0$ would be a clear signature
of a tensor gluonium state.

To summarize, the main conclusion from our study is that the experimental results on 
the $\gamma^\ast\gamma \to f_2(1270)$ transition form factors reported in Ref.~\cite{Masuda:2015yoh} 
appear to be in a very good agreement with QCD scaling predictions starting already at moderate 
$Q^2 \simeq 5\,\text{GeV}^2$. 
The absolute normalization for all helicity form factors 
can be reproduced assuming a 10--15\% lower value of the tensor meson coupling to the 
quark energy-momentum tensor as compared to the estimates existing in the literature, which is well within 
the uncertainty.
These findings are in contrast to the transition form factors
to pseudoscalar $\pi,\eta,\eta'$ mesons where large scaling violations 
have been observed~\cite{Aubert:2009mc,BABAR:2011ad,Uehara:2012ag}.    
If confirmed by future higher-statistics measurements that can come from BELLE~II, 
perfect scaling behavior can be an indication that higher-twist 
and soft corrections are less of an issue for tensor as compared to pseudoscalar mesons. 
This can be interesting in context of the studies of heavy meson 
decays~\cite{Wang:2010ni,Yang:2010qd,Cheng:2010yd,Li:2010ra,Lu:2011jm,Zou:2012sy}
where the effective hard scale is not very large and estimates of preasymptotic corrections are difficult.
In turn, the QCD description implemented in
our analysis can still be improved in many ways, e.g., taking into account deviation from ideal $SU(3)$-flavor 
mixing at hadronic scales, two-loop scale dependence of the couplings, higher-twist and end-point corrections 
to $T_1(Q^2)$, more elaborate models for the DAs, etc. 
%The corrections can be of order 1--3\% for each effect. 
The corresponding studies will become necessary if the accuracy of the experimental data is increased.

%%%%%%%%%%%%%%%%%%%%%%%%%%%%%%%%%%%%%%%%%%%%%%%%%%%%%%%%%%%%%%%%%%%%%%%%%%%%%%%%%%%%%%%%%%%%%%%%%%%%%%%%%%%%%%%%%%%%%%
\section*{Acknowledgments}
%%%%%%%%%%%%%%%%%%%%%%%%%%%%%%%%%%%%%%%%%%%%%%%%%%%%%%%%%%%%%%%%%%%%%%%%%%%%%%%%%%%%%%%%%%%%%%%%%%%%%%%%%%%%%%%%%%%%%%
\addcontentsline{toc}{section}{Acknowledgments}

N.K. is grateful to M. Vanderhaeghen for useful discussions. 
The work by M.S. is supported by a stipend through the F+E (Research and Development) grant by the GSI 
and the Helmholz Graduate School (HGS-HIRe), project number RSCH{\"A}F1416.

\appendix

%%%%%%%%%%%%%%%%%%%%%%%%%%%%%%%%%%%%%%%%%%%%%%%%%%%%%%%%%%%%%%%%%%%%%%%%%%%%%%%%%%%%%%%%%%%%%%%%%%%%%%%%%%%%%%%%%%%%%%
%%%%%%%%%%%%%%%%%%%%%%%%%%%%%%%%%%%%%%%%%%%%%%%%%%%%%%%%%%%%%%%%%%%%%%%%%%%%%%%%%%%%%%%%%%%%%%%%%%%%%%%%%%%%%%%%%%%%%%
%%%%%%%%%%%%%%%%%%%%%%%%%%%%%%%%%%%%%%%%%%%%%%%%%%%%%%%%%%%%%%%%%%%%%%%%%%%%%%%%%%%%%%%%%%%%%%%%%%%%%%%%%%%%%%%%%%%%%%
%%%%%%%%%%%%%%%%%%%%%%%%%%%%%%%%%%%%%%%%%%%%%%%%%%%%%%%%%%%%%%%%%%%%%%%%%%%%%%%%%%%%%%%%%%%%%%%%%%%%%%%%%%%%%%%%%%%%%%

\section*{Appendices}
\addcontentsline{toc}{section}{Appendices}

\renewcommand{\theequation}{\Alph{section}.\arabic{equation}}
\renewcommand{\thetable}{\Alph{table}}
\setcounter{section}{0}
\setcounter{table}{0}

%%%%%%%%%%%%%%%%%%%%%%%%%%%%%%%%%%%%%%%%%%%%%%%%%%%%%%%%%%%%%%%%%%%%%%%%%%%%%%%%%%%%%%%%%%%%%%%%%%%%%%%%%%%%%%%%%%%%%%
\section{Other conventions}\label{app:HelAmp}
%%%%%%%%%%%%%%%%%%%%%%%%%%%%%%%%%%%%%%%%%%%%%%%%%%%%%%%%%%%%%%%%%%%%%%%%%%%%%%%%%%%%%%%%%%%%%%%%%%%%%%%%%%%%%%%%%%%%%%

%\section{The relations between  the  FFs $F_{i}$ measured in
%Ref.\cite{Masuda:2015yoh} and  the form factors  $T_{i}$ defined in Eq.(2.4)}

The experimental  results in Ref.~\cite{Masuda:2015yoh} are presented for a different
set of transition form factors $F_{i}(Q^2)$ suggested  in~\cite{Schuler:1997yw}. 
The form factors  $T_{i}(Q^{2})$ defined in \eqref{defT} are more convenient for the QCD study
but in order to compare our results with the data  we need to establish the precise correspondence 
between these two descriptions. 

In Ref.~\cite{Schuler:1997yw}, the cross section $\sigma_{\lambda_{1}\lambda_{2}}$ for the  production of $f_{2}(1270)$ 
by photons with helicities $\lambda_{1}$ and $\lambda_{2}$ is written as
\begin{equation}
\sigma_{\lambda_{1}\lambda_{2}}=\delta(s-m^{2})8\pi^{2}\frac{5\Gamma_{\gamma\gamma}}{m}~f_{\lambda_{1}\lambda_{2}}(Q^{2}),
\label{sigmaAB}
\end{equation}
where  $s=(q_{1}+q_{2})^{2}$ and $\Gamma_{\gamma\gamma}$ denotes the two-photon
decay width~\eqref{twophotonwidth}. The form factors are defined in terms of the helicity 
cross sections as~\cite{Schuler:1997yw}
\begin{equation}
F_{0}(Q^{2})=\sqrt{\frac{f_{TT}^{\pm\pm}(Q^{2})}{(1+Q^{2}/m^{2})}},
\qquad F_{1}(Q^{2})=\sqrt{\frac{f_{LT}(Q^{2})}{(1+Q^{2}/m^{2})}},
\qquad F_{2}(Q^{2})=\sqrt{\frac{f_{TT}^{\pm\mp}(Q^{2})}{(1+Q^{2}/m^{2})}}.
~\label{Fi:def}
\end{equation} 
Calculation of the helicity cross sections (\ref{sigmaAB}) in terms of the Lorentz
covariant amplitudes similar to  $T_{i}$  was done in
Ref.~\cite{Pascalutsa:2012pr}, see Appendix~C3. Using the expressions presented there we obtain
\begin{eqnarray}
\sigma_{TT}^{\pm\pm}&=&
\delta(s-m^{2})~8\pi^{2}\frac{5\Gamma_{\gamma\gamma}}{m}~\biggl\{
\frac{\Gamma_{\gamma\gamma}^{\Lambda=0}}{\Gamma_{\gamma\gamma}}\left(
1+\frac{Q^{2}}{m^{2}}\right)^{-1}
\left\vert \frac{T_{0}(Q^{2})}{T_{0}(0)}\right\vert ^{2}\biggr\},
\\
\sigma_{LT}&=&\delta(s-m^{2})~8\pi^{2}~\frac{5\Gamma_{\gamma\gamma}}{m}~\biggl\{
\frac{\pi\alpha^{2}}{5m\Gamma_{\gamma\gamma}}
\frac{Q^{2}/m^{2}}{(1+Q^{2}/m^{2})^{3}}\left\vert T_{1}(Q^{2})\right\vert ^{2}\biggr\},
\\
\sigma^{\pm\mp}_{TT}&=&\delta(s-m^{2})~8\pi^{2}\frac{5\Gamma_{\gamma\gamma}}{m}~\biggl\{
\frac{\Gamma_{\gamma\gamma}^{\Lambda=2}}{\Gamma_{\gamma\gamma}}\left(
1+\frac{Q^{2}}{m^{2}}\right)^{-1}
\left\vert \frac{T_{2}(Q^{2})}{T_{2}(0)}\right\vert ^{2}\biggr\},
\end{eqnarray}
where $\Gamma_{\gamma\gamma}^{\Lambda}$ stands for the two-photon decay width of
$f_{2}(1270)$ with the polarization $\Lambda$:
\begin{align}
\Gamma_{\gamma\gamma}^{\Lambda=2}=\frac{\pi\alpha^{2}}{5m}\left\vert T_{2}(0)\right\vert ^{2},
&&
\Gamma_{\gamma\gamma}^{\Lambda=0}=\frac{\pi
\alpha^{2}}{5m}\frac{2}{3}\left\vert T_{0}(0)\right\vert^{2}.
\end{align}
Using these expressions and the definitions in \eqref{Fi:def} one finds
\begin{eqnarray}
F_{0}(Q^{2})&=&\sqrt{\frac{\Gamma_{\gamma\gamma}^{\Lambda=0}}{\Gamma
_{\gamma\gamma}}}\left(  1+\frac{Q^{2}}{m^{2}}\right)  ^{-1}
\left\vert\frac{T_{0}(Q^{2})}{T_{0}(0)}\right\vert ,
\label{def:F0}\\
F_{1}(Q^{2})&=&\sqrt{\frac{\pi\alpha^{2}}{5m\Gamma_{\gamma\gamma}}}\frac
{\sqrt{Q^{2}/m^{2}}}{(1+Q^{2}/m^{2})^{2}}\left\vert T_{1}(Q^{2})\right\vert,
\label{def:F1}\\
F_{2}(Q^{2})&=&\sqrt{\frac{\Gamma_{\gamma\gamma}^{\Lambda=2}}{\Gamma
_{\gamma\gamma}}}\left(  1+\frac{Q^{2}}{m^{2}}\right)  ^{-1}\left\vert
\frac{T_{2}(Q^{2})}{T_{2}(0)}\right\vert\,.
\label{def:F2}
\end{eqnarray}
Experimentally the ratio of the decay widths with $\Lambda=0$ and $\Lambda=2$ is small~\cite{Uehara:2008ep}:
\begin{align}
\frac{\Gamma_{\gamma\gamma}^{\Lambda=0}}{\Gamma_{\gamma\gamma}^{\Lambda=2}}\simeq (3.7\pm 0.3)\times 10^{-2}.
\end{align}
Hence the expressions in (\ref{def:F0}-\ref{def:F2}) can be simplified 
neglecting the contribution of $\Gamma_{\gamma\gamma}^{\Lambda=0}$ in the full decay width:
\begin{eqnarray}
F_{0}(Q^{2})&\simeq&\sqrt{\frac{2}{3}}\left(  1+\frac{Q^{2}}{m^{2}}\right)
^{-1}\left\vert \frac{T_{0}(Q^{2})}{T_{2}(0)}\right\vert ,
\label{F0:T0}\\
F_{1}(Q^{2})&\simeq&\frac{\sqrt{Q^{2}/m^{2}}}{(1+Q^{2}/m^{2})^{2}}\left\vert
\frac{T_{1}(Q^{2})}{T_{2}(0)}\right\vert ,
\label{F1:T1}\\
F_{2}(Q^{2})&\simeq&\left(  1+\frac{Q^{2}}{m^{2}}\right)  ^{-1}\left\vert
\frac{T_{2}(Q^{2})}{T_{2}(0)}\right\vert .
\label{F2:T2}
\end{eqnarray}
We use these simplified relations in order to present the data~\cite{Masuda:2015yoh} in terms of
the $T_i$ form factors that are more suitable for comparison with QCD predictions. 

The effective form factor $F_{f_2}(Q^2)$ is defined in~\cite{Masuda:2015yoh} as
\begin{align}
   F_{f_2}(Q^2) &= \sqrt{F_{0}^2(Q^{2}) + F_{1}^2(Q^{2}) + F_{2}^2(Q^{2})}\,.
\end{align}
It is written in our notation as 
\begin{align}
    (1+Q^2/m^2) F_{f_2}(Q^2) &= 
\sqrt{\frac23 \left\vert\frac{T_{0}(Q^{2})}{T_{2}(0)}\right\vert^2 
     + \frac{Q^2m^2}{(m^2+Q^2)^2} \left\vert\frac{T_{1}(Q^{2})}{T_{2}(0)}\right\vert^2 
     + \left\vert\frac{T_{2}(Q^{2})}{T_{2}(0)}\right\vert^2}.    
\end{align}

For completeness we quote the phenomenological ansatz for the
form factors $F_{i}$ suggested in~\cite{Schuler:1997yw}:
\begin{equation}
F_{0}=(1+Q^{2}/m^{2})^{-2}\frac{1}{\sqrt{6}}\frac{Q^{2}}{m^{2}},~~\ F_{1}%
=(1+Q^{2}/m^{2})^{-2}\frac{Q}{m},~\ F_{2}=(1+Q^{2}/m^{2})^{-2}.
\end{equation}
Note that the asymptotic behavior for the FF $F_{2}$ is different
from the QCD result, see Eq.~\eqref{QCDT2}, because the contribution  of the gluon transversity distribution 
has not been taken into account. More model predictions can be found in
Refs.~\cite{Pascalutsa:2012pr, Achasov:2015pha}.

%%%%%%%%%%%%%%%%%%%%%%%%%%%%%%%%%%%%%%%%%%%%%%%%%%%%%%%%%%%%%%%%%%%%%%%%%%%%%%%%%%%%%%%%%%%%%%%%%%%%%%%%%%%%%%%%%%%%%%
\section{Scale dependence}\label{app:scale}
%%%%%%%%%%%%%%%%%%%%%%%%%%%%%%%%%%%%%%%%%%%%%%%%%%%%%%%%%%%%%%%%%%%%%%%%%%%%%%%%%%%%%%%%%%%%%%%%%%%%%%%%%%%%%%%%%%%%%%

In this Appendix we summarize the scale dependence and mixing under renormalization
to the leading one-loop accuracy for all relevant parameters. In what follows
\begin{align}
  L = \frac{\alpha_s(\mu)}{\alpha_s(\mu_0)}\,, && \beta_0 = \frac{11}{3}N_c - \frac23 n_f\,.  
\end{align}
As already mentioned in the main text, for simplicity, we make use of  
the decoupling scheme, or fixed flavor number scheme (FFNS), such that 
the DAs only involve the three light flavors and  the charm
$c$-quark contributions are included in the coefficient function.
Going over to the variable flavor number scheme (VFNS) is straightforward
but has very limited numerical impact so that we do not implement it in this study.

For definiteness we also assume ideal quark mixing at a low normalization point $\mu_0 = 1$~GeV, 
$f_2 \sim (u\bar u + d\bar d)/\sqrt{2}$. Thus all matrix elements involving strange
quark vanish at this scale, but appear at higher scales because of the evolution.
Staying within the fixed three-flavor scheme we decompose the $SU(2)$-flavor singlet coupling $f_q$ 
in the $SU(3)$-flavor singlet and octet parts that have different scale dependence:
\begin{align}
 f_{(8)} =  \frac{1}{\sqrt{6}}\big( f_u + f_d -2 f_s\big)\,, && f_{(1)} = \frac{1}{\sqrt{3}}\big( f_u + f_d + f_s\big)\,,
\end{align}
where $f_{u,d,s}$ are the couplings for the separate flavors.
Thus
\begin{align}
     f_q (\mu) &= \phantom{-}\sqrt{\frac13}f_{(8)}(\mu) + \sqrt{\frac23} f_{(1)}(\mu)\,, 
\notag\\
     f_s(\mu)  &= - \sqrt{\frac23}f_{(8)}(\mu) + \sqrt{\frac13} f_{(1)}(\mu)\,. 
\end{align}
Ideal mixing at the reference scale implies  
\begin{align}
    f_{(8)}(\mu_0) = \sqrt{\frac13} f_q(\mu_0)\,, && f_{(1)}(\mu_0) =  \sqrt{\frac23} f_q(\mu_0)\,, && f_s(\mu_0) =0 \,. 
\end{align}
The relevant renormalization group equation reads~\cite{Gross:1973ju,Georgi:1951sr,Gross:1974cs}    
\begin{align}
\left( \mu \frac{\partial}{\partial \mu} + \beta(g) \frac{\partial}{\partial g}\right)
\begin{pmatrix}
   f_{(8)} \\
   f_{(1)} \\
   f_g^s
\end{pmatrix}
= 
\frac{\alpha_s}{2\pi}
\begin{pmatrix}
\frac83 C_F && 0 && 0 \\
0 && \frac83 C_F  && -\frac43 \sqrt{n_f} \\
0 && -\frac43 \sqrt{n_f}C_F  && \frac23 n_f
\end{pmatrix}
\begin{pmatrix}
f_{(8)}  \\
f_{(1)} \\
f_g^s
\end{pmatrix},
\label{qgmixcoupling}
\end{align}  
where from one finds
\begin{align}
   f_{(8)}(\mu) &= L^{(\frac{8}{3}C_F)/\beta_0} f_{(8)}(\mu_0)\,,
\notag\\
  f_{(1)}(\mu) &= f_{(1)}(\mu_0) + 
\Big[L^{(\frac{8}{3}C_F+\frac23 n_f)/\beta_0}-1\Big] \left[ \frac{4C_F}{4C_F+n_f}f_{(1)}(\mu_0) - \frac{2\sqrt{n_f}}{4C_F+n_f} f_g^s(\mu_0)\right],
\notag\\
   f_g^s(\mu)  &=  f_g^s(\mu_0) - 
\Big[L^{(\frac{8}{3}C_F+\frac23 n_f)/\beta_0}-1\Big] \left[ \frac{2C_F \sqrt{n_f}}{4C_F+n_f}f_{(1)}(\mu_0) - \frac{n_f}{4C_F+n_f} f_g^s(\mu_0)\right],
\notag\\
   f_g^T (\mu) &=  L^{( \frac{7}{3}C_A + \frac{2}{3}n_f)/\beta_0} f_g^T (\mu_0)\,. 
\end{align}   
The last expression is based on the calculation of the relevant anomalous dimension by  
Hoodbhoy and Ji~\cite{Hoodbhoy:1998vm}.
Note that the following combination of the quark and gluon couplings is scale-independent:
\begin{align}
   \sqrt{n_f}f_{(1)}(\mu) + 2 f_g^s(\mu) = \sqrt{n_f}f_{(1)}(\mu_0) + 2 f_g^s(\mu_0)\,, 
\end{align}
as it corresponds to the matrix element of a conserved current: the traceless part of the QCD energy-momentum tensor.

The scale dependence of the flavor-nonsinglet twist-three couplings $\zeta_3$, $\omega_3$ and $\widetilde \omega_3$ can be 
found, e.g., in~\cite{Ball:1998sk,Ball:2007rt}. Since the twist-three gluon DAs are completely unknown,
using flavor-singlet evolution equations is not justified, and also the numerical difference between flavor-singlet
and flavor-nonsinglet evolution is negligible as compared with the errors on the parameters. 
Staying with the flavor-nonsinglet evolution one obtains 
\begin{align}
 \zeta_3 (\mu) = L^{3(C_A -C_F)/\beta_0} \zeta_3 (\mu_0)\,. 
\end{align} 
The remaining couplings $\omega_3$ and $\widetilde \omega_3$ mix with each other:
\begin{align}
\begin{pmatrix}
  \widetilde \omega_3 \\ \omega_3 
\end{pmatrix}
 (\mu) 
 & = 
L^{\Gamma/\beta_0}  
\begin{pmatrix}
  \widetilde \omega_3 \\  \omega_3 
\end{pmatrix}
(\mu_0)
\end{align}
with the anomalous dimension matrix 
\begin{align}
\Gamma  =
\begin{pmatrix}
\frac{13}{6}C_A -\frac{1}{12} C_F && \frac72 C_A -\frac{21}{4} C_F \\
\frac{1}{6} C_A - \frac14 C_F && \frac{25}{6} C_A - \frac{29}{12} C_F
\end{pmatrix}
~=~
\begin{pmatrix}
\frac{115}{18} && \frac72 \\ \frac{1}{6}    && \frac{167}{18}
\end{pmatrix}.
\end{align}

%%%%%%%%%%%%%%%%%%%%%%%%%%%%%%%%%%%%%%%%%%%%%%%%%%%%%%%%%%%%%%%%%%%%%%%%%%%%%%%%%%%%%%%%%%%%%%%%%%%%%%%%%%%%%%%%%%%%%%
\section{QCD sum rules}\label{app:QCDSR}
%%%%%%%%%%%%%%%%%%%%%%%%%%%%%%%%%%%%%%%%%%%%%%%%%%%%%%%%%%%%%%%%%%%%%%%%%%%%%%%%%%%%%%%%%%%%%%%%%%%%%%%%%%%%%%%%%%%%%%

The twist-three quark-gluon couplings can be estimated from the tensor meson contribution 
to the correlation functions of 
\begin{align}\label{J}
J_{\mu\nu}(x) = \frac12 
\bar q(x) \left[\gamma_\mu i\der_\nu + \gamma_\nu i\der_\mu \right] q(x)\,,  \qquad q\bar q \equiv (u\bar u + d \bar d)/\sqrt{2}\,,
\end{align}
$\der = \derright - \derleft$, 
and the quark-gluon light-ray operators that enter the definition of the corresponding DAs,
\begin{align}
 \mathcal{G}_\alpha(z_1,z_2,z_3;x) &=  \bar q(z_3n+x) ig G_{\alpha n }(z_2n+x) \slashed{n} q(z_1n+x)\,,
\notag\\
 \widetilde{\mathcal{G}}_\alpha(z_1,z_2,z_3;x) &=  \bar q(z_3n+x) g \widetilde{G}_{\alpha n }(z_2n+x) \slashed{n}\gamma_5 q(z_1n+x)\,.
\end{align} 
In particular we consider the following correlation functions:
\begin{align}
  T_{\alpha n n ,\bar n\bar n} &= i \int \!d^4x\,  e^{iqx}  \langle 0| T\{J_{\bar n\bar n}(x) \mathcal{G}_\alpha(z_1,z_2,z_3;0)\}|0\rangle
\notag\\ &= 
\Big[ \bar n_\alpha (qn)-q_\alpha(n\bar n)\Big] (n\bar n)  \int\mathcal{D}\alpha\, e^{iqn\sum z_k \alpha_k}\, T(q^2;\alpha) + \mathcal{O}(n_\alpha)
%\end{align}
\\
%\begin{align}
  \widetilde{T}_{\alpha n n ,\bar n\bar n} &= i \int \!d^4x\,  e^{iqx}  \langle 0| T\{J_{\bar n\bar n}(x) \widetilde{\mathcal{G}}_\alpha(z_1,z_2,z_3;0)\}|0\rangle
\notag\\ &= 
\Big[ \bar n_\alpha (qn)-q_\alpha(n\bar n)\Big] (n\bar n)  \int\mathcal{D}\alpha\, e^{iqn\sum z_k \alpha_k}\, \widetilde{T}(q^2;\alpha) + \mathcal{O}(n_\alpha)\,,
\end{align}
where it is assumed that the auxiliary light-like vectors are chosen such that
\begin{align}
 \qquad (q\bar n) = 0\,, \qquad (q n) \slashed{=} 0\,.
\end{align}
We obtain
\begin{align}
 T(q^2;\alpha) &= \frac{\alpha_s}{2\pi^3}\frac{\Gamma[2-d]}{[-q^2]^{2-d}}
  \alpha_1\alpha_2\alpha_3 \left[\frac{1-2\alpha_1}{1-\alpha_1}+ \frac{1-2\alpha_3}{1-\alpha_3}+ 4 \right]
+\frac{\langle g^2 G^2\rangle}{12\pi^2} \frac{\Gamma[2-\dhalf]}{[-q^2]^{2-\dhalf}}  \alpha_1\alpha_3 \delta(\alpha_2) 
\notag\\&\quad
 + \frac{2}{9} g^2\langle \bar q q \rangle^2 \frac{1}{-q^2} \delta(\alpha_1)\delta(\alpha_3)\,,
\notag\\
\widetilde{T}(q^2;\alpha) &=  \frac{\alpha_s}{2\pi^3}\frac{\Gamma[2-d]}{[-q^2]^{2-d}}
  \alpha_1\alpha_2\alpha_3 \left[\frac{1-2\alpha_1}{1-\alpha_1} - \frac{1-2\alpha_3}{1-\alpha_3}\right]
+ 0\cdot \langle g^2 G^2\rangle + 0 \cdot \langle \bar q q \rangle^2\,,
\end{align}
where $\langle g^2 G^2\rangle$ is the gluon condensate and $\langle \bar q q \rangle$ is the quark condensate and we used
the usual factorization approximation for the vacuum expectation values of the four-fermion operators.
Note that the correlation function $\widetilde{T}(q^2;\alpha)$ does not receive nonperturbative corrections 
(to this power accuracy in the OPE and to the leading order in the strong coupling).

The contribution of $f_2(1270)$ to these correlation functions is
\begin{align} 
 g_{\alpha\alpha'}^\perp T_{\alpha' n n ,\bar n\bar n} &
= - q_{\alpha}^\perp (n\bar n)^2\frac{|f_q|^2 m^4}{m^2-q^2}\int\mathcal{D}\alpha\,e^{iqn\sum \alpha_k z_k} \Phi_{3}(\alpha) +\ldots
\end{align}
and similar for $\widetilde{T}_{\alpha' n n ,\bar n\bar n}$, so that taking moments and applying the Borel transformation one ends up with 
the sum rules
\begin{align}
    |f_q|^2 m^4\, e^{-m^2/M^2} &= \frac{3}{40\pi^2} \int_0^{s_0} s^2 ds\, e^{-s/M^2}
  -\frac29 \Big\langle \frac{\alpha_s}{\pi} G^2\Big\rangle \int_0^{s_0} ds\, e^{-s/M^2}
 + \frac{16\pi\alpha_s}{9}\langle\bar q q \rangle^2\,,
\notag\\
    |f_q|^2  m^4  e^{-m^2/M^2} \zeta_{3} & =  
\frac{7\alpha_s}{720\pi^3} \int_0^{s_0}\! s^2 ds\, e^{-s/M^2}
+ \frac{1}{18}\Big\langle \frac{\alpha_s}{\pi} G^2\Big\rangle \int_0^{s_0}\!ds\, e^{-s/M^2}
+ \frac{8\pi\alpha_s}{9}\langle \bar q q \rangle^2, 
\notag\\
    |f_q|^2  m^4  e^{-m^2/M^2} \frac34 \omega_{3} &=   
- \frac{7\alpha_s}{1440\pi^3} \int_0^{s_0}\! s^2 ds\, e^{-s/M^2} 
- \frac{1}{6}\Big\langle \frac{\alpha_s}{\pi} G^2\Big\rangle\int_0^{s_0}\!ds\, e^{-s/M^2} 
+ \frac{32\pi\alpha_s}{9}\langle \bar q q \rangle^2, 
\notag\\
    |f_q|^2  m^4  e^{-m^2/M^2}\frac{1}{28} \widetilde{\omega}_{3} &= \frac{\alpha_s}{1440\pi^3}\int_0^{s_0}\! s^2 ds\, e^{-s/M^2}    
\end{align}
where, for completeness, we added in the first line the sum rule for the coupling $|f_q|^2$ derived in \cite{Aliev:1981ju,Aliev:1982ab} and 
reanalyzed more recently in \cite{Cheng:2010hn}. 
Using the value $s_0= 2.53$~GeV$^2$~\cite{Cheng:2010hn} and the interval $1.0 < M^2 <1.4$~GeV$^2$ 
for the Borel parameter we obtain from this sum rule for the standard values of the gluon $\Big\langle \frac{\alpha_s}{\pi} G^2\Big\rangle = 0.012$~GeV$^4$ and
quark $\langle \bar q q \rangle = (-240\,\text{MeV})^3$ condensates
\begin{align}
   f_q(\mu = 1\,\text{GeV}) = 101(10)~\text{MeV}\,.
 \end{align}
The quoted error corresponds to a 50\% uncertainty in the gluon condensate, other uncertainties are much smaller. 
The quark-gluon couplings $\zeta_3,\omega_3,\widetilde{\omega}_3$ can best be estimated 
by taking the ratios of the corresponding sum rules to the sum rule for  $|f_q|^2$. 
Using the same values of input parameters we obtain
\begin{align}
   \zeta_3 = 0.15(8)\,,\qquad \omega_3 = -0.2(3) \qquad \widetilde{\omega}_3 = 0.06(1)\,.
%\ll \zeta_3,\omega_3
\end{align} 
The given values correspond to the scale 1 GeV. Note that the uncertainty in $\omega_3$ is very large because of the 
cancellations between gluon and quartic condensates. For $\widetilde{\omega}_3$ the leading nonperturbative corrections vanish 
and the perturbative contribution is very small. It is tempting to conclude that $\widetilde{\omega}_3$ is  much smaller 
than  $\zeta_3$ and $\omega_3$, but the number given above should be viewed with caution as the sum rule for this  coupling is likely 
to be dominated by uncalculated higher-order corrections and/or condensates of higher dimension.  

Estimates of gluon couplings are notoriously very difficult, see e.g.~\cite{Novikov:1981xi}.
A limited insight can be obtained by considering the correlation function 
\begin{align}
  \mathbb{G}_{\mu\nu} &= i \int \!d^4x\,  e^{iqx}  \langle 0| T\{G^a_{n\mu}(x)G^a_{n\nu}(x)G^b_{\bar n\xi}(0)G^b_{\bar n\xi}(0)\}|0\rangle
\notag\\
&=  (q_\mu q_\nu-\frac12 q^2 g_{\mu \nu}) (n\bar n )^2  \mathbb{G}_1(q^2) + \frac12 g_{\mu\nu}  (n\bar n )^2  \mathbb{G}_2(q^2) + \ldots   
\end{align}
where the ellipses stand for the structures $\sim n_\mu,\bar n_\mu, n_\nu,\bar n_\nu$ and, as above, we assumed that $(\bar n q)=0$.
Since tensor $2^{++}$ gluonium (glueball) states are expected to be rather heavy, see e.g.~\cite{Gregory:2012hu},
by choosing a sufficiently low interval of duality in these invariant functions one can constrain the contribution of $f_2(1270)$. 
The leading contributions to the invariant functions $\mathbb{G}_1(q^2)$ and $\mathbb{G}_2(q^2)$ are, retaining singular terms only
(cf.~\cite{Novikov:1981xi}), 
\begin{align}
 \mathbb{G}_1(Q^2) ~=~ \frac{\langle G^2\rangle}{3 q^2}\,,  
\qquad
 \mathbb{G}_2(Q^2) ~=~  \frac{1}{5\pi^2} \frac{\Gamma[-\dhalf]}{[-q^2]^{-\dhalf}} + 0\cdot \langle G^2\rangle\,, 
\end{align}
and the contribution of the tensor $f_2(1270)$ meson is
\begin{align}
 \mathbb{G}_1(Q^2) ~=~  - \frac{f^S_g  f^T_g m^2 }{m^2 - q^2} +\ldots\,,
\qquad
 \mathbb{G}_2(Q^2) ~=~  \frac{|f^S_g|^2 m^4}{m^2 - q^2} +\ldots\,,
\end{align}
respectively. Thus
\begin{align}
    |f_g^S|^2 m^4  e^{-m^2/M^2} ~\approx~  \frac{1}{10\pi^2} \int_0^{s_0} s^2 ds\, e^{-s/M^2}\,,
\qquad
    f_g^S f_g^T m^2  e^{-m^2/M^2} ~\approx~    \frac13 \langle G^2\rangle\,. 
\end{align}
Taken at face value, these sum rules suggest that both couplings are of the order of 100~MeV (which 
should be viewed as an estimate from above), and have the same sign. 

\begin{figure}
\centerline{\includegraphics[width=0.35\linewidth]{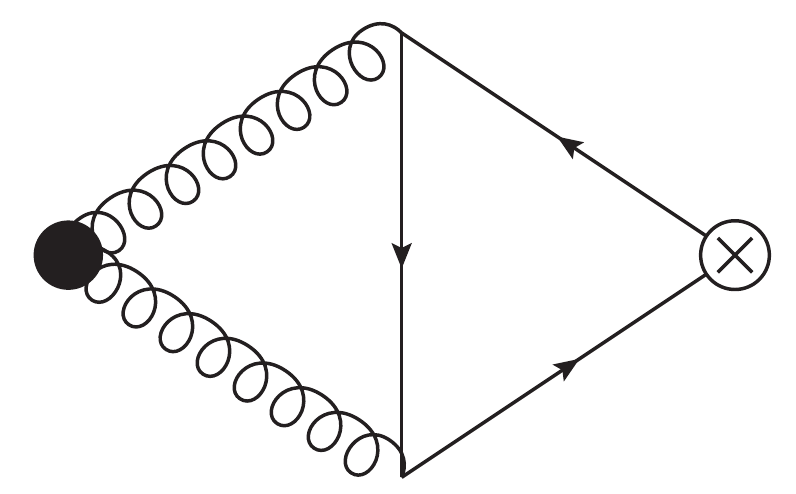}}
\caption{\sf The leading contribution to the correlation function in Eq.~\eqref{quarkgluoncorrelator}.}
\label{fig:QuarkGluonSR}
\end{figure}

A somewhat better estimate can be obtained by considering the correlation function
\begin{align}
  \mathbb{H}_{\mu\nu} &= i \int \!d^4x\,  e^{iqx}  \langle 0| T\{G^a_{n\mu}(x)G^a_{n\nu}(x)J_{\bar n\bar n} (0)\}|0\rangle
\notag\\
&=  (q_\mu q_\nu-\frac12 q^2 g_{\mu \nu}) (n\bar n )^2  \mathbb{H}_1(q^2) + \frac12 g_{\mu\nu}  (n\bar n )^2  \mathbb{H}_2(q^2) + \ldots   
\label{quarkgluoncorrelator}
\end{align}
Assuming $(q \bar n)=0$, the contribution of $f_2(1270)$ to this correlator is 
\begin{align}
 \mathbb{H}_1(Q^2) ~=~  \frac{f_q  f^T_g m^2 }{m^2 - q^2} +\ldots\,,
\qquad
 \mathbb{H}_2(Q^2) ~=~  - \frac{ f_q f^S_g m^4}{m^2 - q^2} +\ldots\,.
\end{align}
The leading contribution in QCD is given by the Feynman diagram shown in Fig~\ref{fig:QuarkGluonSR}.
%%%%%%%%%%%%%%%%%%%%%%%%%%%%%%%%%%%%%%%%%%%%%%%%%%%%%%%%%%%%%%%%%%%%%%%%%%%%%%%%%%%%%%%%%%%%%%%%%%%%%%%%%%%%%%%%%%%%%%
%%%%%%%%%%%%%%%%%%%%%%%%%%%%%%%%%%%%%%%%%%%%%%%%%%%%%%%%%%%%%%%%%%%%%%%%%%%%%%%%%%%%%%%%%%%%%%%%%%%%%%%%%%%%%%%%%%%%%%
We obtain
\begin{eqnarray}
\mathbb{H}_1&=&\sqrt{2}\frac{\alpha_s}{(4\pi)^3}q^2 C_AC_F\left[\frac{8}{9}\ln\Big(\frac{\mu^2}{-q^2}\Big)+\frac{139}{54}\right] +\ldots
\nonumber\\
\mathbb{H}_2&=& - \sqrt{2}\frac{\alpha_s}{(4\pi)^3}q^4 C_AC_F\left[\frac{8}{15}\ln^2\Big(\frac{\mu^2}{-q^2}\Big)
   +\frac{598}{225}\ln\Big(\frac{\mu^2}{-q^2}\Big)+\frac{5627}{1500}\right] +\ldots
\end{eqnarray}
where from one obtains the sum rules 
\begin{eqnarray}
 f_q f_g^T m^2 e^{-m^2/M^2} &\approx& \frac{8\sqrt{2}}{9} \frac{\alpha_s}{(4\pi)^3}C_A C_F \int_0^{s_0}ds \, s e^{-s/M^2},
\notag\\
 f_q f_g^S m^4 e^{-m^2/M^2} &\approx&  \frac{\sqrt{2}\alpha_s}{(4\pi)^3}C_A C_F \int_0^{s_0}ds \, s^2  e^{-s/M^2}
\Big[ \frac{16}{15} \ln \frac{\mu^2}{s} + \frac{598}{225}\Big].
\end{eqnarray}
Dividing these expressions by the sum rule for $|f_q|^2$ we obtain for the same values of parameters
\begin{align}
  f_g^T/f_q = 0.25-0.29\,, &&  f_g^S/f_q = 0.53-0.58\,.  
\label{gluoncouplingsfromQCDSR} 
\end{align}
Again, it appears that the two gluon couplings have the same sign. 
The accuracy of this calculation is very difficult to quantify, 
we view the numbers in \eqref{gluoncouplingsfromQCDSR} as order-of-magnitude estimates only.

%%%%%%%%%%%%%%%%%%%%%%%%%%%%%%%%%%%%%%%%%%%%%%%%%%%%%%%%%%%%%%%%%%%%%%%%%%%%%%%%%%%%%%%%%%%%%%%%%%%%%%%%%%%%%%%%%%%%%%%%%
\section{$f_{g}^{S}$ from the radiative decay $\Upsilon (1S)\rightarrow \protect\gamma f_{2}$}
\label{app:Upsilon}
%%%%%%%%%%%%%%%%%%%%%%%%%%%%%%%%%%%%%%%%%%%%%%%%%%%%%%%%%%%%%%%%%%%%%%%%%%%%%%%%%%%%%%%%%%%%%%%%%%%%%%%%%%%%%%%%%%%%%%%%%

%
%%%%%%%%%%%%%%%%%%%%%%%%%%%%%%%%%%%%%%%%%%%%%%%%%%%%%%%%%%%%%%%%%%%%%%%%%%%%%%%%%%%%%%%%%%%%%%%%%%%%%%%%%%%%%%%%%%%%%%
\begin{figure}[t]
\centerline{\includegraphics[width=0.35\linewidth]{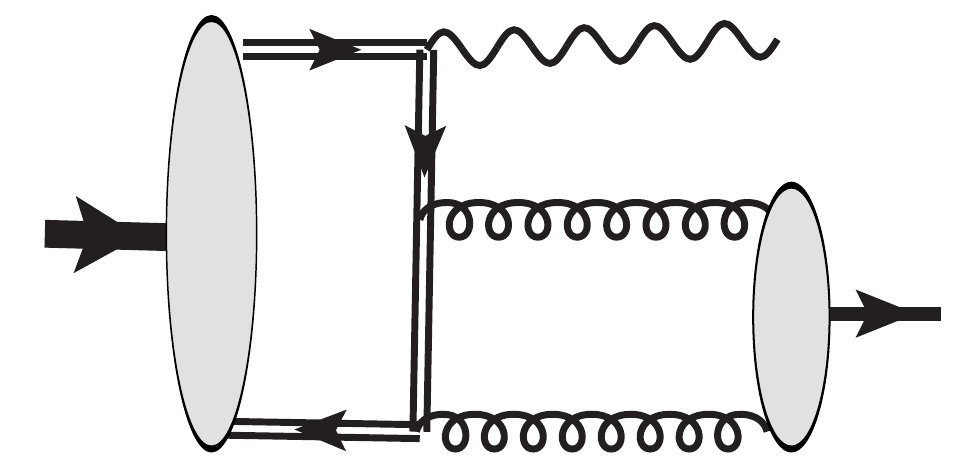}}
\caption{\sf The leading contribution to the radiative decay $\Upsilon (1S)\rightarrow \gamma f_{2}(1270)$. }
\label{fig:jpsi2f2gamma}
\end{figure}
%%%%%%%%%%%%%%%%%%%%%%%%%%%%%%%%%%%%%%%%%%%%%%%%%%%%%%%%%%%%%%%%%%%%%%%%%%%%%%%%%%%%%%%%%%%%%%%%%%%%%%%%%%%%%%%%%%%%%%
%

The scalar gluon coupling $f_{g}^{S}$ can be estimated from the
bottomonium decay $\Upsilon (1S)\rightarrow \gamma f_{2}(1270)$. 
The calculation was already discussed in Ref.~\cite{Fleming:2004hc}
where it was shown that the dominant contribution comes from the two-quark  $Q\bar{Q}$ component of the
bottomonium wave function; the contribution of higher Fock states is
suppressed by the small relative velocity of the heavy quarks. To the
leading-order accuracy the decay amplitude is described by the diagram in Fig.~\ref{fig:jpsi2f2gamma}. 
The result reads 
\begin{equation}
A\left[ \Upsilon (1S)\rightarrow \gamma\,f_{2}\right] =(\epsilon _{\gamma
}^{\ast }\cdot \epsilon _{\Upsilon }^{\ast })\sqrt{2M_{\Upsilon }}\sqrt{%
\frac{3}{2\pi }}\frac{R_{10}(0)}{m_{b}^{4}}2\pi \alpha
_{s}ee_{b}~e_{nn}^{(\lambda )}~f_{g}^{S}m_{f}^{2}
\frac14 \int_0^1 \frac{du}{u\bar u}\, \phi_{g}^{S}(u)\,,
%\int_{-1}^{1}dt~\frac{\phi_{g}^{S}(t)}{1-t^{2}},
\end{equation}%
where $\epsilon _{\gamma }^{\ast }$ and $\epsilon _{\Upsilon }^{\ast }$ are
the polarization vectors of the photon and heavy meson, respectively, $m_b$ is the $b$-quark (pole) mass and $R_{10}(0)$ denotes
the radial wave function of $\Upsilon (1S)$ at the origin. Potentially there could be also a contribution of the 
transverse DA $\phi _{g}^{T}(t)$, but the corresponding terms cancel to the leading-order accuracy. 

In order to avoid the dependence on the nonperturbative parameter $R_{10}(0)$ it is
convenient to consider the ratio 
\begin{equation}
\frac{Br[\Upsilon (1S)\rightarrow \gamma ~f_{2}]}{Br[\Upsilon(1S)\rightarrow e^{+}e^{-}]}
=
\frac{64\pi}{3}\frac{\alpha_{s}^{2}(4m_{b}^{2})}{\alpha}
\left(1- \frac{m^{2}}{M_{\Upsilon }^{2}} \right)
%~\frac{M_{\Upsilon }^{2}-m^{2}}{M_{\Upsilon }^{2}}
\frac{\left[ f_{g}^{S}\,I_{g}^{S}\right] ^{2}}{m_{b}^{2}}\,,
\label{RBr}
\end{equation}
where this dependence cancels.
Here we used the notation $I_{g}^{S}$ for the integral  
\begin{equation}
~I_{g}^{S}(\mu )= \frac14 \int_0^1 \frac{du}{u\bar u}\, \phi_{g}^{S}(u,\mu)\,.
%\int_{-1}^{1}dt~\frac{\phi _{g}^{S}(t,\mu )}{1-t^{2}}.
\end{equation}
For the asymptotic DA $\phi_{g}^{S}(u,\mu) = 30 u^2(1-u)^2$ one obtains $I_{g}^{S}=\frac54$.
The branching fractions on the l.h.s. of Eq.(\ref{RBr}) are known, see~\cite{Agashe:2014kda}:
\begin{align}
&Br[\Upsilon (1S)\rightarrow \gamma ~f_{2}]=(1.01\pm 0.09)\times 10^{-4}, 
\notag\\ 
&Br[\Upsilon(1S)\rightarrow e^{+}e^{-}]= (2.38\pm 0.11)\times 10^{-2}\,.
\end{align}%
Using $m_{b}\simeq 4.8$GeV, $\alpha _{s}(4m_{b}^{2})=0.176$
%$\alpha _{s}(4m_{b}^{2})=0.215$, 
and $\alpha \simeq 1/137$  we obtain 
\begin{equation}
|f_{g}^{S}\,I_{g}^{S}|(\mu^2 =4m_{b}^{2})= (18.6\pm 1.9)\,,\text{MeV}\,,
%18.6\,\text{MeV}\,,
\end{equation}
where from, for the asymptotic DA, 
%\begin{equation}
%I_{g}^{S}(\mu =4m_{b}^{2})=\int_{-1}^{1}dt~\frac{\phi _{g}^{S}(t,4m_{b}^{2})%
%}{1-t^{2}}\simeq \frac{5}{4},
%\end{equation}%
one finds
\begin{equation}
f_{g}^{S}(\mu^2 = 4m_{b}^{2})=(14.9\pm 0.8)\,\text{MeV.}
\label{fgs}
\end{equation}
Here we tacitly assumed that this coupling is positive (with respect to $f_q$), as suggested 
by the QCD sum rule analysis in Appendix~\ref{app:QCDSR}. 
The given error bar reflects experimental uncertainties only. The theoretical uncertainties
are much larger so that we estimate the overall accuracy of the value in \eqref{fgs} as 30--50\%.
 
This result appears to support an intuitive picture that the gluon coupling of 
``ordinary'' quark-antiquark mesons 
is very small at hadronic scales and is generated entirely by the evolution.
Indeed, assuming  $f_{q}(1\,\text{GeV})=101(10)$\,MeV and $f^s_{g}(1\,\text{GeV}$)=0 and 
using the expressions collected in Appendix~\ref{app:scale} one finds
\begin{equation}
f_{g}^{S}(\mu^2 = 4m_{b}^{2})=(25\pm 3)\,\text{MeV.}
\end{equation}
This number is in a reasonable agreement with the above extraction from the bottomonium radiative decay
having in mind the theoretical uncertainties.

%
%
%%%%%%%%%%%%%%%%%%%%%%%%%%%%%%%%%%%%%%%%%%%%%%%%%%%%%%%%%%%%%%%%%%%%%%%%%
%%%%%%%%%%%%%%%%%%%%%%%%%%%%%%%%%%%%%%%%%%%%%%%%%%%%%%%%%%%%%%%%%%%%%%%%%
%

\end{document}